\documentclass[final,1p,times,twocolumn]{elsarticle}
\usepackage{graphicx}
\usepackage{amssymb}
\usepackage{soul} % PARA TACHADO \textst
\usepackage{geometry} % Optional: for better page layout if needed
\usepackage{tabularx}
\usepackage{amsmath} % For \boldsymbol if needed, though \Sigma is standard
\usepackage{float}
\usepackage[utf8]{inputenc}
\usepackage{xcolor}
\usepackage{subcaption}
\usepackage{booktabs}
\usepackage{multirow}
\usepackage{rotating}
\usepackage{makecell}
\usepackage[hyphens]{url}
\usepackage{multirow}
\usepackage{arydshln}  % for dashed lines. ASC june 2021
\usepackage{mathtools}
\usepackage{array}
\newcolumntype{H}{>{\setbox0=\hbox\bgroup}c<{\egroup}@{}}
\usepackage{bm}
\let\<\textless
\let\>\textgreater
\let\|\textbar
\usepackage{algorithm} 
\usepackage{algpseudocode,algorithmicx} % added in iGNGSVM
\algnewcommand{\algorithmicand}{\textbf{ and }}
\algnewcommand{\algorithmicor}{\textbf{ or }}
\algnewcommand{\OR}{\algorithmicor}
\algnewcommand{\AND}{\algorithmicand}
\algnewcommand{\var}{\texttt}

\algnewcommand{\IIf}[1]{\State\algorithmicif\ #1\ \algorithmicthen}
\algnewcommand{\EndIIf}{\unskip\ \algorithmicend\ \algorithmicif}

\journal{arXiv}

\begin{document}

\begin{frontmatter}


\title{ProteuS: A Generative Approach for Simulating Concept Drift in Financial Markets}
% \title{Generating Financial-like Synthetic Regime Changes for High-frequency Data Streams}
% Generation of Financial-like Semi-synthetic Data Streams



%\address[label2]{Address Two\fnref{label4}}
%\fntext[label3]{I also want to inform about\ldots}
%\fntext[label4]{Small city}

\author[label3,label1]{Andr\'es L. Su\'arez-Cetrulo\corref{cor1}}
\address[label3]{Ireland's Centre for Artificial Intelligence (CeADAR), School of Computer Science, University College Dublin, D04 V2N9 Dublin, 
(Ireland)}
\ead{andres.suarez-cetrulo@ucd.ie}
\cortext[cor1]{Corresponding author}

\author[label2]{Alejandro Cervantes}
% \ead{acervant@inf.uc3m.es}
\ead{alejandro.cervantesrovira@unir.net}

\address[label2]{Escuela Superior de Ingeniería y Tecnología, Universidad Internacional de La Rioja (UNIR), Logroño (Spain)}

\author[label1]{David Quintana}%\fnref{label1}}
\ead{dquintan@inf.uc3m.es}

\address[label1]{Department of Computer Science and Engineering, Universidad Carlos III de Madrid, Avda. Universidad 30, 28911 Legan\'es (Spain).}

\begin{abstract}
    Financial markets are complex, non-stationary systems where the underlying data distributions can shift over time, a phenomenon known as regime changes, as well as concept drift in the machine learning literature. These shifts, often triggered by major economic events, pose a significant challenge for traditional statistical and machine learning models. A fundamental problem in developing and validating adaptive algorithms is the lack of a ground truth in real-world financial data, making it difficult to evaluate a model's ability to detect and recover from these drifts. This paper addresses this challenge by introducing a novel framework, named ProteuS, for generating semi-synthetic financial time series with pre-defined structural breaks. Our methodology involves fitting ARMA-GARCH models to real-world ETF data to capture distinct market regimes, and then simulating realistic, gradual, and abrupt transitions between them. The resulting datasets, which include a comprehensive set of technical indicators, provide a controlled environment with a known ground truth of regime changes. An analysis of the generated data confirms the complexity of the task, revealing significant overlap between the different market states. We aim to provide the research community with a tool for the rigorous evaluation of concept drift detection and adaptation mechanisms, paving the way for more robust financial forecasting models. ProteuS is available in our public GitHub repository: \url{https://github.com/cetrulin/regime-switching-series-generator}
\end{abstract}


\end{frontmatter}

% \linenumbers

\section{Introduction}
\label{S:introduction}

The inherent nature of financial markets is that of a complex, adaptive system, marked by high volumes of noisy, unstructured data where underlying relationships are often obscured \citep{Abu-Mostafa1996IntroductionForecasting, Lo2004, Huang2005ForecastingMachine}. While conventional forecasting in finance has long relied on traditional statistical models, these methods are frequently constrained by an assumption of linearity in the data-generating process; this is a significant drawback given the market's intrinsic complexity \citep{cavalcante_computational_2016}. In contrast, machine learning (ML) approaches have emerged as a powerful alternative, demonstrating a remarkable capacity to model nonlinear dynamics without pre-specified assumptions, often yielding results that have challenged established financial expertise \citep{Hsu2016BridgingEconomists, Atsalakis2009SurveyingMethods}.

Within the field of computational finance, the phenomenon of concept drift, defined as shifts in the statistical properties of the data stream and the relationships between model variables over time, has gained significant importance \citep{Tsymbal2004TheWork, Suarez-Cetrulo2019}. Major economic events, including global pandemics, financial crises, and speculative bubbles in assets like cryptocurrencies, have underscored the non-stationary character of markets and their susceptibility to abrupt structural transformations \citep{Ardia2019}. Such episodes of instability are often accompanied by swift and significant alterations in key financial metrics, including mean returns, volatility, and cross-asset correlations \citep{ang2012regime}. This has led to a growing body of research suggesting that markets operate in distinct regimes or states, which can shift, evolve, or reappear over time in response to macroeconomic forces such as inflation, deflation, and changes in supply and demand dynamics \citep{Munnix2012IdentifyingMarket, Tsang2020, Hu2015ConceptMarket, Gu2016OnlineStream}.

In ML research for data stream learning, to effectively handle concept drift, researchers must determine when a true change occurs and how long it persists. However, a significant challenge in real-world data is the absence of a ground truth, that is, a definitive record of when these concept changes in fact happen. Without this ground truth, it is difficult to accurately evaluate the performance of algorithms designed to detect and adapt to these shifts. Furthermore, there is no absolute baseline for what a classifier's predictive accuracy should be if a drift does not occur, making it infeasible to estimate whether a classifier has successfully recovered from a drift.

To address this evaluation challenge, this paper introduces the ProteuS framework, a novel methodology for generating semi-synthetic financial time series with pre-defined structural breaks. The main contribution of this paper is this framework, which creates datasets with properties similar to those of stock market prices, inspired by relevant approaches from the literature \citep{anderson2019recurring, Shaker2015}. By using these datasets, we establish a known ground truth for when concept changes occur. In the future, these datasets will be pre-processed and fed as a data stream into our proposed algorithm for price trend classification. The algorithm's results can then be rigorously compared to the ground truth, allowing for a precise evaluation of its ability to recognize concept drifts and manage its concept history through successful retrievals and insertions. This paper details the generation of these crucial synthetic datasets, which form the foundation for a robust evaluation of algorithms operating in dynamic financial environments.

The remainder of this paper is structured as follows. Section \ref{sec:literature} reviews related work on concept drift and financial time series. Section \ref{sec:approach} presents our data generation framework in detail. Section \ref{sec:expsetup} outlines the experimental setup used to validate our approach. Section \ref{sec:results} discusses the results of our experiments and their limitations. Finally, Section \ref{sec:conclusion} concludes the paper and suggests directions for future research.

\section{Literature Review}
\label{sec:literature}

\subsection{Concept Drifts in Financial Markets}

The theoretical basis for market predictability is a subject of ongoing debate. The traditional Efficient Market Hypothesis (EMH) suggests that asset prices fully reflect all available information, making markets inherently unpredictable \citep{fama_efficient_1970}. However, a more contemporary view is offered by the Adaptive Market Hypothesis (AMH), which posits that market efficiency is not static but evolves over time \citep{Lo2004}. According to the AMH, market participants use heuristics and adapt to changing market conditions, leading to periods where market dynamics can be predicted. This perspective aligns with empirical evidence showing that market predictability varies depending on the specific time horizon and prevailing conditions \citep{URQUHART201639}.

This dynamic nature of markets, as described by the AMH, is central to our research. In computational finance, the observable shifts in market behavior are referred to by various terms, including regime changes \citep{ang2012regime, Hammerschmid2018, Guidolin2006, baek2020covid}, structural breaks \citep{andreou2009structural, Pettenuzzo2011, de2018advances}, or distinct market states \citep{Munnix2012IdentifyingMarket}. These terms all describe a similar phenomenon: long periods of relative stability that are punctuated by abrupt and significant changes in the market's underlying dynamics \citep{Hammerschmid2018}.

From an academic perspective, this phenomenon is analogous to the problem of concept drift, which is defined as a change in the underlying data distribution and the relationship between a model's input features and its target variable over time \citep{Tsymbal2004TheWork}. The relevance of concept drift to finance is not merely theoretical; studies have shown that major economic events, like the Great Recession, manifest as detectable concept drifts in the generative processes of financial data \citep{masegosa2020analyzing}. Researchers have successfully identified these regime changes by analyzing shifts in key financial metrics. For instance, \cite{Tsang2020} developed a framework to distinguish between market regimes based on the volatility and velocity of price returns, while \cite{Munnix2012IdentifyingMarket} identified different market states by observing changes in the correlation structure among assets in the S\&P 500.

While the existence of these market regimes and structural breaks is well-established, the primary obstacle for developing and validating algorithms that can adapt to them is the absence of a definitive ground truth in real-world data. It is impossible to know with certainty when a true regime change began or ended, making it exceedingly difficult to measure an algorithm's performance in detecting and adapting to these shifts. This evaluation challenge has been a significant bottleneck in the field \citep{cavalcante_computational_2016}. To properly assess an algorithm's ability to handle concept drift, a controlled environment is needed where structural breaks are known and can be precisely controlled. This necessity motivates the core contribution of our work: a framework for generating synthetic financial data with verifiable regime changes.

\subsection{Financial-like Data Generation}

A prerequisite for generating realistic financial data is the ability to first identify and model the distinct market regimes present in historical data. Unsupervised learning methods provide a powerful toolkit for this task by discovering latent structures and representing different market states. A prominent approach is model-based clustering, particularly using the Expectation-Maximization (EM) algorithm to fit a mixture of Gaussian distributions to the data, thereby representing each regime as a distinct statistical model \citep{dellaert2002expectation}. The Baum–Welch algorithm, a specific variant of EM, has been effectively used with Hidden Markov Models (HMMs) to detect change points and model market states \citep{Tsang2020, dias2015clustering, kritzman2012regime}. In parallel, econometric time series models, such as Autoregressive Moving Average with Generalized Autoregressive Conditional Heteroskedasticity (ARMA-GARCH) models, are widely employed to capture the specific volatility and return dynamics that characterize different market states \citep{cetrulo2024machine}.

An alternative strategy involves data partitioning methods that summarize the data distribution into a set of locally optimal structures \cite{Angelov2017}. Prototype generation techniques, such as Learning Vector Quantization (LVQ), Self-Organizing Maps (SOM), and Growing Neural Gas (GNG) \citep{xu2012incremental, oisvm, fritzke1996growing}, have been successfully applied in the financial domain for data partitioning and model selection \citep{choudhury_real_2014-1, pavlidis2006financial}. 

Once distinct market regimes are identified and modeled, the next challenge is to simulate the transitions between them. Simply generating data from static, isolated models is insufficient; a realistic simulation must also capture the dynamics of the shift from one state to another. As highlighted in surveys on the topic, evaluating algorithms designed for changing environments necessitates datasets with known drift points, for which synthetic data generation is a must-do \citep{SUAREZCETRULO2023118934}. These transitions can vary significantly in their nature, ranging from sudden or abrupt shifts that occur over a short period to more gradual or incremental changes that unfold over a longer duration.

The work by Shaker and Hüllermeier provides an approach for simulating these dynamics \citep{Shaker2015}. Although their concept of ``recovery analysis" was initially developed to evaluate a learner's performance after a change, its implementation relies on the generation of semi-synthetic datasets with controlled, known transitions between different generative processes. Their framework models the transition between two concepts as a gradual shift, where the output is a weighted combination of the old and new generative processes. This methodology is crucial for creating data that is not idealized or unrealistic, but rather reflects the complex dynamics of real-world shifts. Our work is directly inspired by this approach, aiming to apply a similar transition-modeling technique to generate financial-like data with verifiable structural breaks, thereby creating the necessary ground truth for robust algorithm evaluation.

% \section{The Proteus Framework for Data Generation}
% \section{Generation of Financial-like Semi-synthetic Data Streams}
\section{The ProteuS Framework for Financial-like Data Generation}
\label{sec:approach}

\begin{figure}[hbt]
    \centering
    \includegraphics[width=0.85\textwidth]{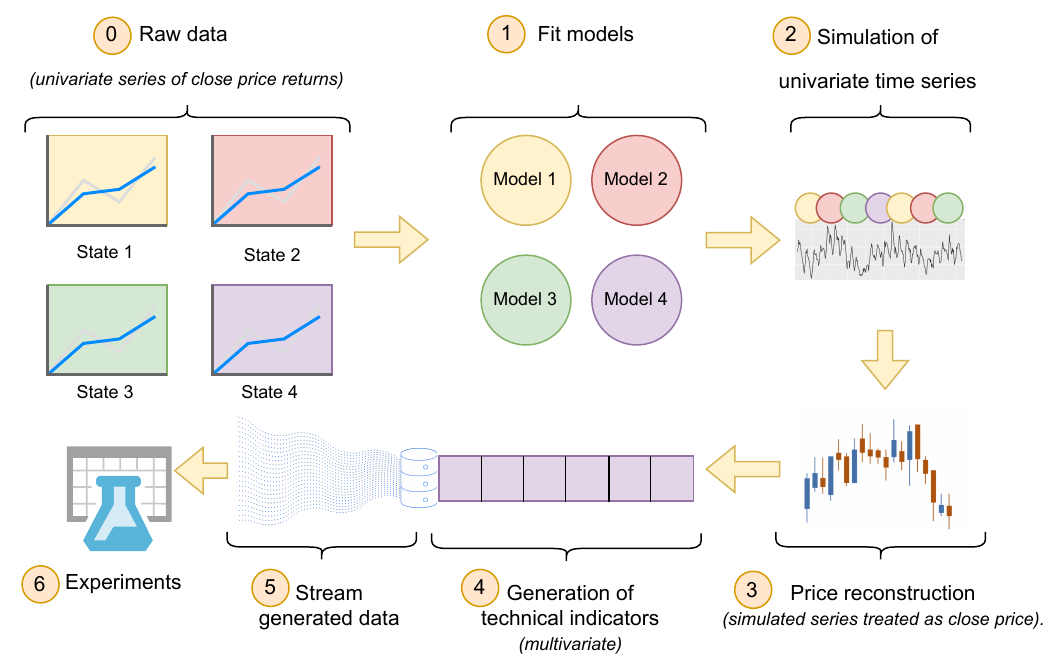}
    % \caption{The end-to-end pipeline of the data generation framework, from raw data selection and model fitting to the final creation of a labeled data stream with technical indicators.}
    \caption{The end-to-end pipeline of the ProteuS framework, from raw data selection and model fitting to the final creation of a labeled data stream with technical indicators.}
    %  https://drive.google.com/file/d/1RZjHTeCuQXQlB2PRWAISs_hJSeU50_L7/view?usp=sharing
    \label{fig:end-to-end}
\end{figure}

A primary challenge in evaluating algorithms designed for dynamic financial markets is the absence of a ground truth for structural breaks. To overcome this limitation, we developed the \textbf{ProteuS} framework for generating semi-synthetic data streams that embed known, controllable regime changes.  
This section details the multi-stage process of the ProteuS framework, which involves modeling distinct market states using econometric models, simulating realistic transitions between these states, and engineering a feature set suitable for machine learning tasks. The complete end-to-end process is illustrated in Figure~\ref{fig:end-to-end} and is covered in the following subsections. 

ProteuS is a three-stage process as follows. Stage 1 of ProteuS fits models to specific market regimes from raw data (steps 1 and 2 in Figure~\ref{fig:end-to-end}); Stage 2 simulates univariate time-series with regime switches (step 3), and Stage 3 reconstructs prices and generates technical indicators as well as target features for later supervised ML models training (steps 3 and 4 in Figure~\ref{fig:end-to-end}).

% \subsection{Modeling Market Regimes with ARMA-GARCH}
% \label{sec:modeling_regimes}
\subsection{Stage 1 (S1): Model Fitting and Optimization} \label{sec:armagarchLags}

The first step of the ProteuS framework is to capture the distinct statistical properties of different market regimes. We hypothesize that these regimes, characterized by unique return and volatility dynamics, can be effectively modeled using established econometric techniques.

For each of the market states selected, we fitted an Autoregressive Moving Average with Generalized Autoregressive Conditional Heteroskedasticity (ARMA-GARCH) model to its price return series. This class of models is well-suited for financial data as it can capture both the mean-reverting behavior of returns (ARMA component) and the time-varying volatility clustering (GARCH component).

To determine the optimal structure for each model, we performed a grid search to select the orders ($p, q$) for both the ARMA and GARCH components. The optimal orders were chosen based on the model that yielded the lowest Akaike Information Criterion (AIC), a standard measure for model selection that balances goodness-of-fit with model complexity \citep{burnham2004multimodel}. The AIC is defined as:
\begin{equation}
    AIC=2k-2ln(L) \label{eq:5AIC}
\end{equation}
where $k$ is the number of estimated parameters in the model and $L$ is the maximized value of the likelihood function.

\subsection{Stage 2 (S2): Simulating Regime Transitions}
\label{sec:simulating_transitions}

With the generative models for each market state established, the next step in ProteuS is to simulate the transitions between them. Our approach is inspired by the work of Shaker and Hüllermeier \cite{Shaker2015}, % on recovery analysis, 
which utilizes a controlled simulation of concept drifts.

We specify a transition map to ProteuS that pre-defines the sequence of regime changes, the starting point of each transition, and its duration. We defined two types of transitions to capture different market dynamics:
\begin{itemize}
    \item \textbf{Abrupt drifts:} A rapid transition completed over 100 data instances.
    \item \textbf{Gradual drifts:} A slower, more incremental transition completed over 1,000 data instances.
\end{itemize}

\begin{figure}[hbt]
    \centering
    \includegraphics[trim={0 0 0 0},clip, width=1.0\linewidth]{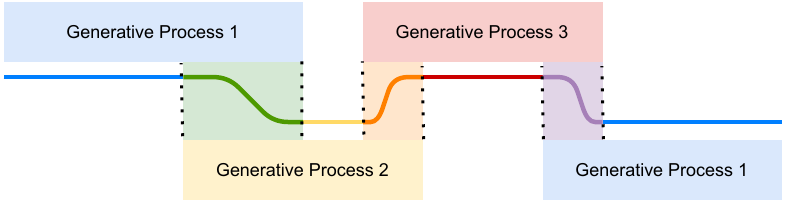}
    \caption[Concept drifts as sigmoidal transitions between generative processes.]{A visual representation of the sigmoidal transition mechanism between generative processes. During a regime change, the simulated output (middle line) is a weighted average of the outputs from the old and new models, ensuring a smooth transition.}
    \label{fig:5generativeprocesses}
\end{figure}

During a transition from an old state to a new state, the generated return at each time step is a weighted average of the outputs from the two corresponding ARMA-GARCH models. The weights are adjusted at each step according to a sigmoidal function, ensuring a smooth and realistic shift from the old regime to the new one. This process is illustrated in Figure~\ref{fig:5generativeprocesses}. The simulation begins by generating data from the first model in the sequence. As the simulation progresses, the output of the current generative process is used as the input for the next time step, creating a continuous and path-dependent time series.

\subsection{Stage 3 (S3): Post-Processing and Feature Engineering}
\label{sec:post_processing}

The final stage of ProteuS transforms the raw simulated return series into a labeled dataset suitable for ML experiments. First, the time series of returns is reconstructed into a price series by cumulatively applying the returns to an initial price. From this reconstructed price series, we compute a comprehensive set of technical indicators, which serve as the feature set for our classification task. The selection of these indicators is based on their prevalence in the financial forecasting literature \citep{Kara2011PredictingExchange}.

Finally, a target for supervised ML training is generated for each time step. In this paper, building on the work in \citep{Suarez-Cetrulo2019}, we prepare the dataset for a price trend classification task. A label of `1' is assigned if the price increases from the previous time step to the current one (uptrend), and a label of `0' is assigned otherwise (stable or downtrend). The result of this entire process is a semi-synthetic data stream with a rich feature set and a known ground truth of when structural breaks occur, providing an ideal environment for the rigorous evaluation of concept drift-aware algorithms.

\section{Experimental design}
\label{sec:expsetup}

% To evaluate the performance of algorithms in detecting and adapting to regime changes rigorously, we have designed a series of experiments based on the semi-synthetic data streams generated by our framework ProteuS. 
This section details the experimental protocol followed to test our framework ProteuS through a series of experiments generating financial-like semi-synthetic data streams. This includes the specification of the ground truth states, the design of the transition map, the data preparation and feature engineering process, and a validation of the inherent separability of the simulated market states. Thus, we elaborate on the experimental setup for Stages 1 to 3 from Section 3 in the following three subsections.

\subsection{S1 - Ground Truth Market States and ARMA-GARCH fitting process}
The core of our experimental design is the generation of a collection of synthetic data streams, constructed from four distinct ground truth market states. These states serve as the building blocks for our simulation, providing a controlled environment where the timing and nature of regime changes are known \textit{a priori}.

To ensure our synthetic data reflects a diverse set of realistic market conditions, we selected four Exchange-Traded Funds (ETFs) as archetypes for our market states. For each ETF, we used the first 1,000 data points of 5-minute resolution price data from January 2020. The selected states are:
\begin{enumerate}
    \item \textbf{State 1 (SPY):} A period of lateral movement with an initial uptrend followed by a slight downtrend, representing a typical equities market behavior.
    \item \textbf{State 2 (PFF):} A consistent uptrend, characteristic of a bullish fixed-income securities market.
    \item \textbf{State 3 (VNQ):} A volatile lateral movement, capturing the fluctuating nature of real estate markets.
    \item \textbf{State 4 (BWX):} A stable lateral movement with a slow uptrend, representing a low-volatility international bond market.
\end{enumerate}

\begin{figure}[hbt]
    \centering
    \subcaptionbox[S1: Equities (SPY)]{S1: Equities (SPY)\label{fig:SPYclose}}
    [0.48\textwidth]{\includegraphics[trim={0cm 1.3cm 0cm 0cm},clip, width=0.48\textwidth]{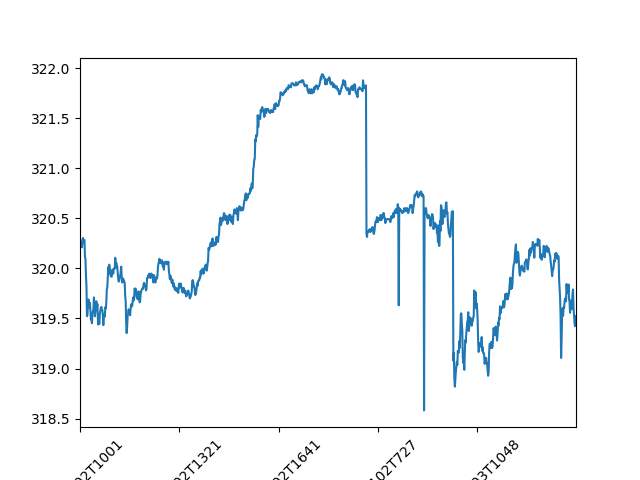}}
    \subcaptionbox[S2 - Fixed-income preferred (PFF)]{S2 - Fixed-income preferred (PFF)\label{fig:PFFclose}}
    [0.48\textwidth]{\includegraphics[trim={0cm 1.3cm 0cm 0cm},clip, width=0.48\textwidth]{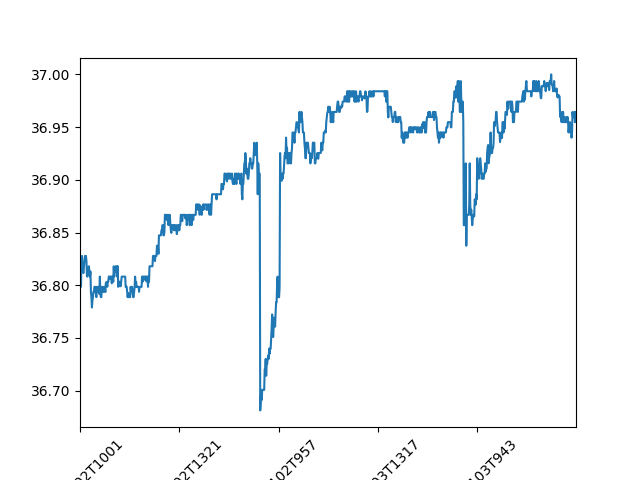}}
    \subcaptionbox[S3 - Real-state (VNQ)]{S3 - Real-state (VNQ)\label{fig:VNQclose}}
    [0.48\textwidth]{\includegraphics[trim={0cm 1.3cm 0cm 0cm},clip, width=0.48\textwidth]{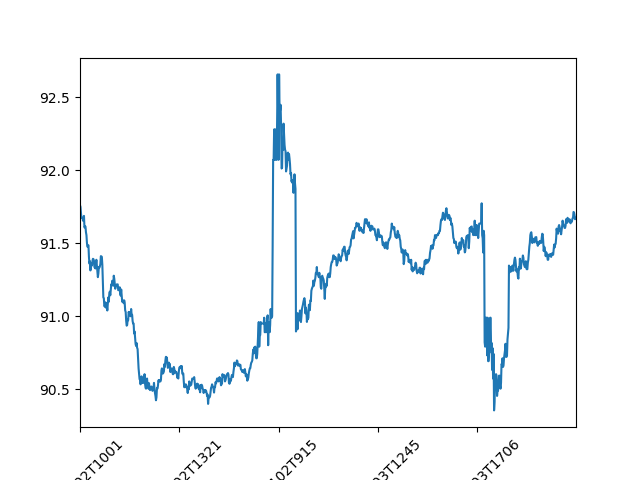}}
    \subcaptionbox[S4 - International bonds (BWX)]{S4 - International bonds (BWX)\label{fig:BWXclose}}
    [0.48\textwidth]{\includegraphics[trim={0cm 1.3cm 0cm 0cm},clip, width=0.48\textwidth]{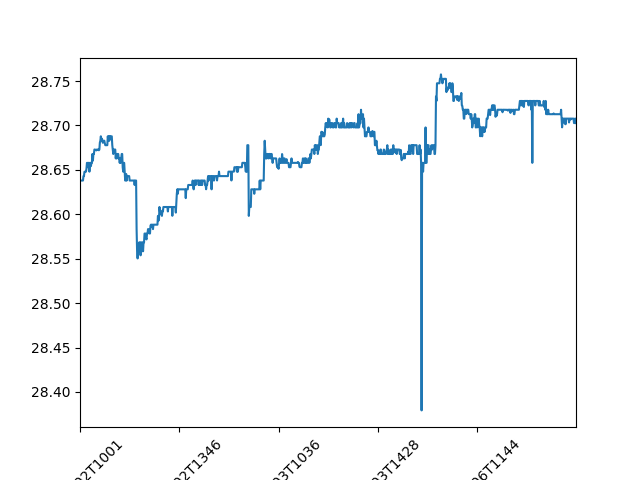}}
    \caption[Series of close prices used for the generative processes.]{The historical close price series for the four selected ETFs, each representing a distinct ground truth market state used to fit the generative ARMA-GARCH models.}
    \label{fig:rawstates}
\end{figure}

The historical price series for these four states is depicted in Figure~\ref{fig:rawstates}. To capture the unique dynamics of each state, we fitted an ARMA-GARCH model to the price return series of each corresponding ETF. %, as detailed in Section~\ref{sec:modeling_regimes}. 
The final optimized orders for each generative model are presented in Table~\ref{fig:5armagarchLags}.

\begin{table}[hbt!]
    \centering
    \small
    \begin{tabular}{lrrrr}
        \toprule
        ETF & $p_{ARMA}$ & $q_{ARMA}$ & $p_{GARCH}$ & $q_{GARCH}$ \\
        \midrule
        SPY & 9 & 25 & 4 & 4 \\
        PFF & 12 & 1 & 1 & 5 \\
        VNQ & 9 & 5 & 1 & 4 \\
        BWX & 21 & 2 & 1 & 1 \\
        \bottomrule
    \end{tabular}
    \caption[Orders of the four ARMA-GARCH models fitted.]{The optimized ARMA and GARCH orders for the four models representing the distinct market states, selected based on the lowest Akaike Information Criterion.}
    \label{fig:5armagarchLags}
\end{table}

\subsection{S2 - Simulation using a Ground Truth Transition Map}

For our experiments, we generated 30 unique data streams using the ProteuS framework, each containing 1.5 million instances. Within each stream, a regime change occurs every 5,000 instances, resulting in a total of 300 ground truth transitions. The transitions between the four states are randomized, with an even distribution of both gradual (1,000 instances) and abrupt (100 instances) drifts. A detailed summary of the transition map, which is consistent across all generated streams, is provided in Table~\ref{tab:transition_map}. For the machine learning experiments, the first 500,000 instances of each stream are reserved for model pre-training and parameter tuning, while the remaining 1 million instances are used for the final prequential evaluation.

\begin{table}[hbt!]
  \resizebox{.8\textwidth}{!}{
  \begin{minipage}{\textwidth}
    \begin{tabular}{lllr}
    \toprule
    From & To & Duration & Starting in instance \# ($k$ represents thousands) \\
    \midrule
     \multirow{5}{*}{1}
     & \multirow{2}{*}{2} & \multirow{2}{*}{100}  &   5k, 245k, 485k, 725k, 965k, 1205k, 1445k, 905k, 185k, 425k, 665k, 1145k, 1385k,  \\
     &                   &                       &   605k, 125k, 365k, 845k, 1085k, 1325k, 65k, 305k, 545k, 1025k, 785k, 1265k \\
     & \multirow{2}{*}{3} & 100  &   1355k, 755k, 155k, 1475k, 875k, 275k, 995k, 395k, 1115k, 515k, 35k, 1235k, 635k \\
     &  & 1k &           1055k, 455k, 1175k, 575k, 1295k, 695k, 1415k, 815k, 215k, 95k, 935k, 335k \\
     & \multirow{2}{*}{4} & 100  &  1245k, 45k, 405k, 285k, 165k, 525k, 645k, 765k, 885k, 1005k, 1125k, 1365k, 1485k \\
     &  & 1k &           105k, 585k, 465k, 225k, 345k, 705k, 825k, 945k, 1065k, 1185k, 1305k, 1425k \\
    \midrule
     \multirow{5}{*}{2}
     & \multirow{2}{*}{1} & 100  &              1290k, 1410k, 90k, 330k, 210k, 570k, 450k, 690k, 810k, 930k, 1050k, 1170k \\
     &  & 1k &         1470k, 30k, 510k, 150k, 390k, 270k, 630k, 750k, 870k, 990k, 1110k, 1230k, 1350k \\
     & \multirow{2}{*}{3} & 100  &              70k, 190k, 550k, 430k, 310k, 670k, 790k, 910k, 1030k, 1150k, 1270k, 1390k \\
     &  & 1k &         1210k, 1330k, 10k, 490k, 370k, 250k, 130k, 610k, 730k, 850k, 970k, 1090k, 1450k \\
     & \multirow{2}{*}{4} & \multirow{2}{*}{1k}  &  1435k, 235k, 415k, 1135k, 655k, 175k, 895k, 1375k, 115k, 835k, 355k, 1075k, \\
     &                   &                       &  595k, 1195k, 1315k, 475k, 535k, 775k, 55k, 295k, 1015k, 1255k, 1495k, 955k, 715k \\
    \midrule
    \multirow{6}{*}{3} 
     & \multirow{2}{*}{1} & 100  &                   100k, 460k, 340k, 220k, 580k, 700k, 820k, 940k, 1060k, 1180k, 1300k, 1420k \\
     &  & 1k &              40k, 160k, 520k, 400k, 280k, 640k, 760k, 880k, 1000k, 1120k, 1360k, 1480k, 1240k \\
     & \multirow{2}{*}{2} & 100  &           25k, 265k, 145k, 385k, 505k, 625k, 745k, 865k, 985k, 1105k, 1225k, 1345k, 1465k \\
     &  & 1k &                   1285k, 85k, 565k, 445k, 325k, 205k, 685k, 805k, 925k, 1045k, 1165k, 1405k \\
     & \multirow{2}{*}{4} & 100  &           1095k, 495k, 1215k, 615k, 1335k, 735k, 135k, 1455k, 855k, 15k, 255k, 975k, 375k \\
     &  & 1k &                   1395k, 795k, 195k, 915k, 315k, 1035k, 435k, 555k, 1155k, 75k, 1275k, 675k \\
    \midrule
     \multirow{6}{*}{4} 
     & \multirow{2}{*}{1} & 100  &           120k, 480k, 360k, 240k, 600k, 720k, 840k, 960k, 1080k, 1200k, 1320k, 1440k \\
     &  & 1k &   60k, 180k, 540k, 300k, 420k, 660k, 780k, 900k, 1020k, 1140k, 1260k, 1380k, 1500k \\
     & \multirow{2}{*}{2} & 100  &           1430k, 110k, 350k, 470k, 230k, 590k, 710k, 830k, 950k, 1070k, 1190k, 1310k \\
     &  & 1k &   1250k, 1370k, 50k, 170k, 530k, 410k, 290k, 650k, 770k, 890k, 1010k, 1130k, 1490k \\
     & \multirow{2}{*}{3} & 100  &              80k, 320k, 200k, 560k, 440k, 680k, 800k, 920k, 1040k, 1160k, 1280k, 1400k \\
     &  & 1k &   20k, 140k, 500k, 380k, 260k, 620k, 740k, 860k, 980k, 1100k, 1220k, 1340k, 1460k \\
    \bottomrule
    \end{tabular}
     \end{minipage}
     }
    \caption[300 ground truth changes in synthetic sets for 1.5M examples.]{The ground truth transition map for the first 1.5 million instances of the synthetic data streams, detailing the timing, destination state, and duration of each of the 300 simulated regime changes.}\label{tab:transition_map}
\end{table}

\subsection{S3 - Data Preparation and Feature Engineering}
\label{sec:feature_eng}

The synthetic price return series generated by our framework is transformed into a feature-rich dataset suitable for a binary classification task. This involves reconstructing the price series and then engineering a set of technical indicators.

\begin{table}[hbt!]
  \centering
  \footnotesize
	\begin{tabular}{ll}
	 \toprule
   Indicator & Formula \\
   \midrule
    A/D  & $\frac{H_t-C_{t-1}}{H_t-L_t}$\\
    CCI  & $\frac{M_t-SM_t}{0.015D_t}$\\
    LWR  & $\frac{H_n-C_t}{H_n-L_n}\times 100$\\
    MACD & $MACD(n)_{t-1}+2/n+1\times(DF_t-MACD(n)_{t-1})$\\
    MOM  & $C_t-C_{t-n}$\\
    RSI  & $100-\frac{100}{1+(\sum^{n-1}_{i=0}Up_{t-i}/n)/(\sum^{n-1}_{i=0}Dw_{t-i}/n)}$ \\
    SMA  & $\frac{C_t+C_{t-1}+...+C_{t-n+1}}{n}$\\
    SD   & $\frac{\sum^{n-1}_{i=0}K_{t-i}\%}{n}$\\
    SK   & $\frac{C_t-LL_{t-n}}{HH_{t-n}-LL_{t-n}}\times 100$\\
    WMA  & $\frac{n \times C_t +(n-1) \times C_{t-1}+...+C_{t-n+1}}{n+(n-1)+...+1}$\\
   \bottomrule
    \multicolumn{2}{l}{\begin{minipage}{5.2in}\baselineskip=12pt \scriptsize $C_t$: closing price; $L_t$: lowest price; $H_t$: highest price at time $t$; $DF$: $EMA(12)_t-EMA(26)_t$; EMA: Exponential moving average; $EMA(k)_t$: $EMA(k)_{t-1}+\alpha \times (c_t-EMA(k)_{t-1})$; $\alpha$: smoothing factor: $2/1+k$; $k$: time period of $k$ minute exponential moving average; $LL_t$ and $HH_t$: mean lowest low and highest high in the last $t$ minutes; $M_t:H_t+L_t+C_t/3$; $SM_t: \sum^n_{i=1}M_{t-i+1})/n$; $D_t:(\sum^n_{i=1}|M_{t-i+1}-SM_t|)/n; Up_t$: upward price change; $Dw_t$: downward price change at time $t$. \\
   \end{minipage}
   } \\
  \end{tabular}
\caption{Technical indicators used and formulas as reported in \citep{Kara2011PredictingExchange}.\label{tbl:3indicators:kara}}
\end{table}

First, the return series is converted into a price series. This is then used to create Open-High-Low-Close (OHLC) data, where the open price is the previous time step's close, and the high and low are set to the close price. From this OHLC data, we compute a set of technical indicators that are commonly used in the financial forecasting literature \citep{Hsu2016BridgingEconomists}. The selection is partially derived from the widely used collection proposed by \cite{Kara2011PredictingExchange}, augmented with additional moving averages to capture trends over multiple time horizons. The formulas for a representative subset of these indicators are provided in Table~\ref{tbl:3indicators:kara}. This comprehensive feature set is designed to capture different aspects of market dynamics, from trend and momentum to volatility. 

The full list is: i) Commodity Channel Index (CCI), ii) Larry William's R (WILLR), iii) Momentum (MOM), iv) Moving average convergence divergence (MACD), v) Relative Strength Index (RSI), vi) Simple n-minute moving average (SMA) over 5 and 10 lags, vii) Stochastic D\% (SD), viii) Stochastic K\% (SK), ix) Weighted n-minute moving average (WMA) over a 10 lag, x) Exponential n-minute moving average (EMA) over a 10 lag, xi) Average directional n-minute moving average (ADW) over a 10 lag, xii) Triangular n-minute moving average (TRIMA) over a 10 lag, xiii) Rate of change (ROC) over a 10-minute lag, xiv) Bollinger upper and lower bands, and xv) Aroon Up/Down.
The set of features utilized in this study partially derives from a widely used collection initially presented by \cite{Kara2011PredictingExchange}. 
We have augmented this base set with the inclusion of additional short-term (5-minute) moving averages. The original set proposed by Kara et al. comprised ten technical indicators selected based on their prominence in publications by domain experts and researchers in financial forecasting; this indicator set has subsequently been used in various related studies \citep{Hsu2016BridgingEconomists}.

\section{Experimental Results and Discussion}
\label{sec:results}

This section presents the results of our experiments to validate our data generation framework (namely ProteuS) and analyzes the resulting datasets. We begin by providing an overview of the 30 generated time series, followed by a statistical analysis of the engineered feature space. Finally, we conduct a validation of the separability of the raw and generated market states to characterize the complexity of the data produced.

\subsection{Overview of the Generated Synthetic Data Streams}
\label{sec:overview_synthetic}

\begin{figure*}[hbt!]
    \centering
    \subcaptionbox[Stream 1]{\label{fig:appendix:synthetic:a}}
    [0.24\textwidth]{\includegraphics[clip, trim=0cm 0.5cm 0.5cm 1.3cm, width=0.99\linewidth]{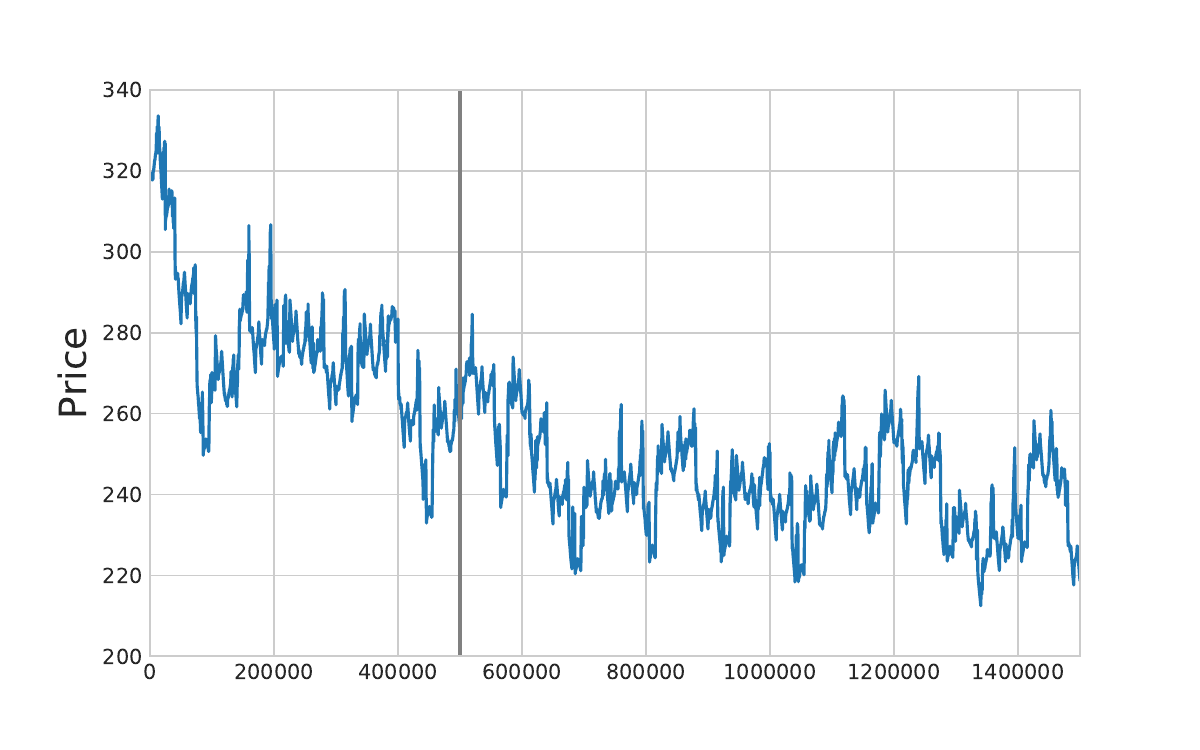}}
    \subcaptionbox[Stream 2]{\label{fig:appendix:synthetic:b}}
    [0.24\textwidth]{\includegraphics[clip, trim=0cm 0.5cm 0.5cm 1.3cm, width=0.99\linewidth]{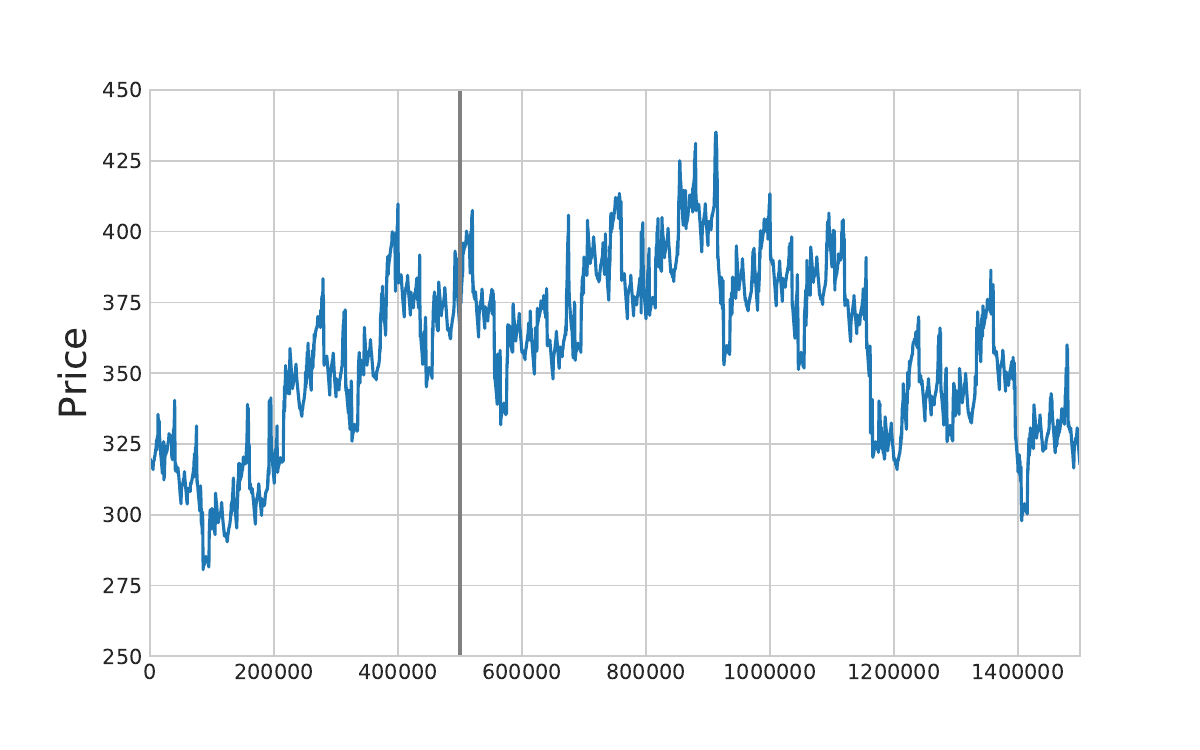}}
    \subcaptionbox[Stream 3]{\label{fig:appendix:synthetic:c}}
    [0.24\textwidth]{\includegraphics[clip, trim=0cm 0.5cm 0.5cm 1.3cm, width=0.99\linewidth]{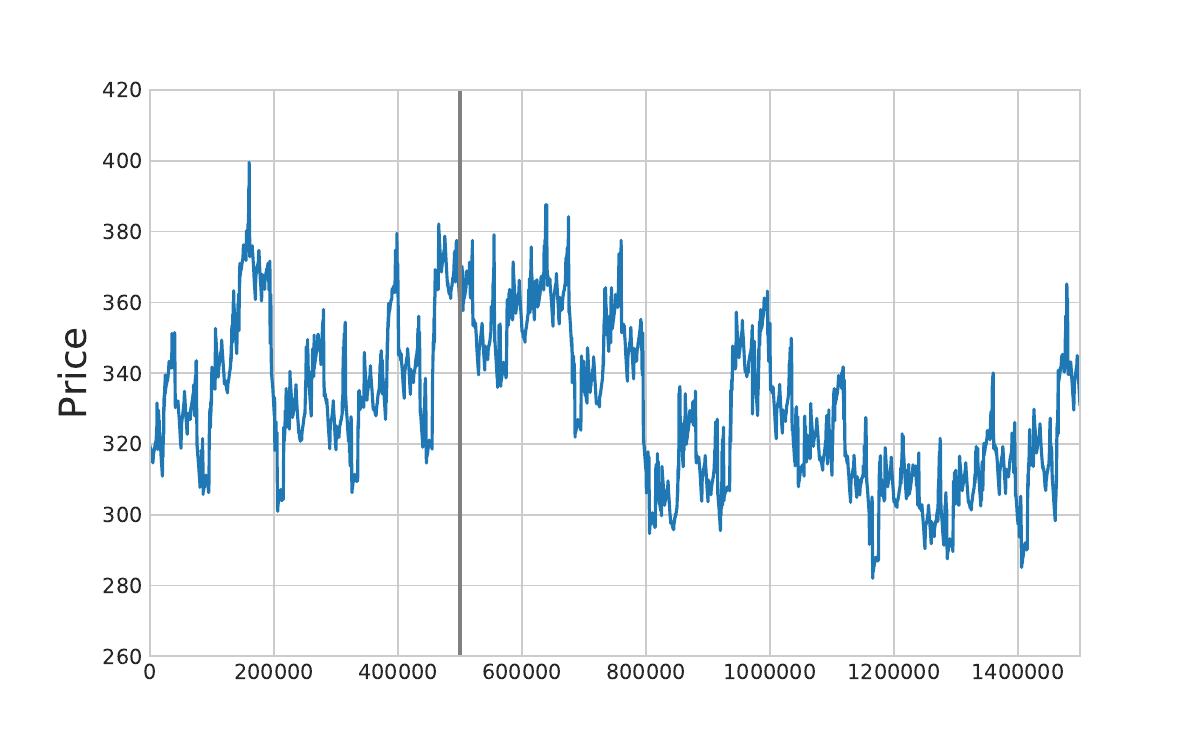}}
    \subcaptionbox[Stream 4]{\label{fig:appendix:synthetic:d}}
    [0.24\textwidth]{\includegraphics[clip, trim=0cm 0.5cm 0.5cm 1.3cm, width=0.99\linewidth]{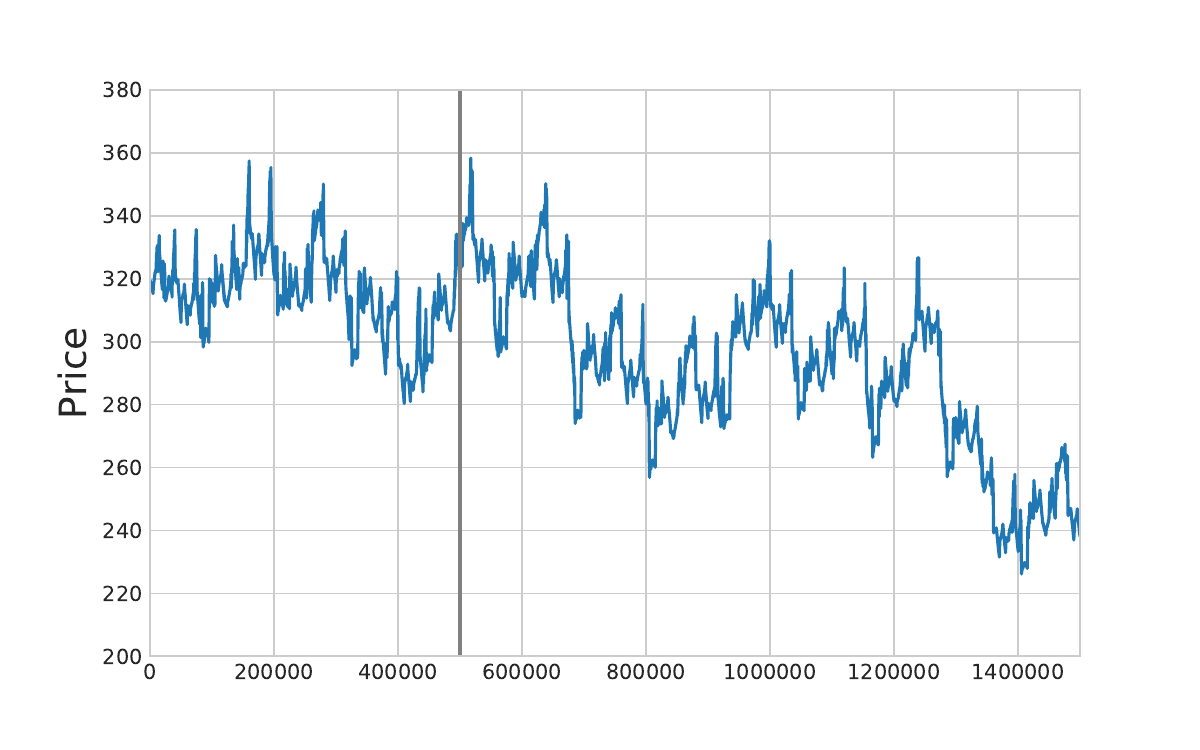}}
    \subcaptionbox[Stream 5]{\label{fig:appendix:synthetic:e}}
    [0.24\textwidth]{\includegraphics[clip, trim=0cm 0.5cm 0.5cm 1.3cm, width=0.99\linewidth]{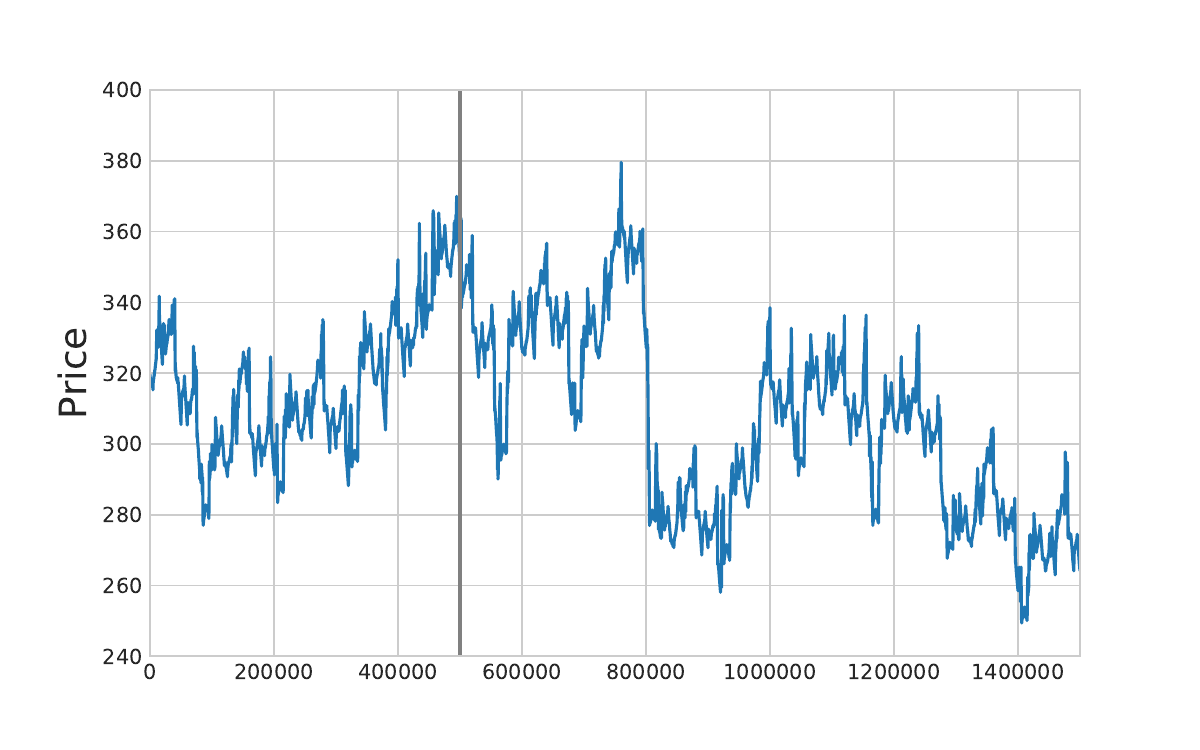}}
    \subcaptionbox[Stream 6]{\label{fig:appendix:synthetic:f}}
    [0.24\textwidth]{\includegraphics[clip, trim=0cm 0.5cm 0.5cm 1.3cm, width=0.99\linewidth]{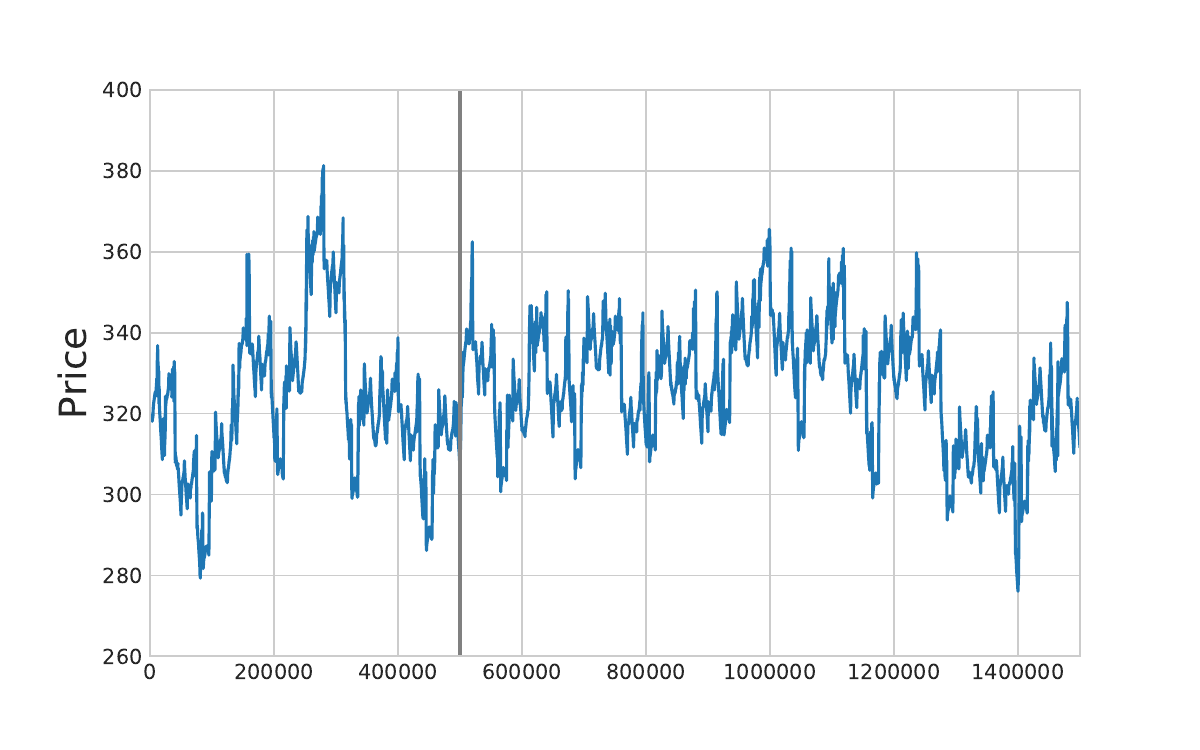}}
    \subcaptionbox[Stream 7]{\label{fig:appendix:synthetic:g}}
    [0.24\textwidth]{\includegraphics[clip, trim=0cm 0.5cm 0.5cm 1.3cm, width=0.99\linewidth]{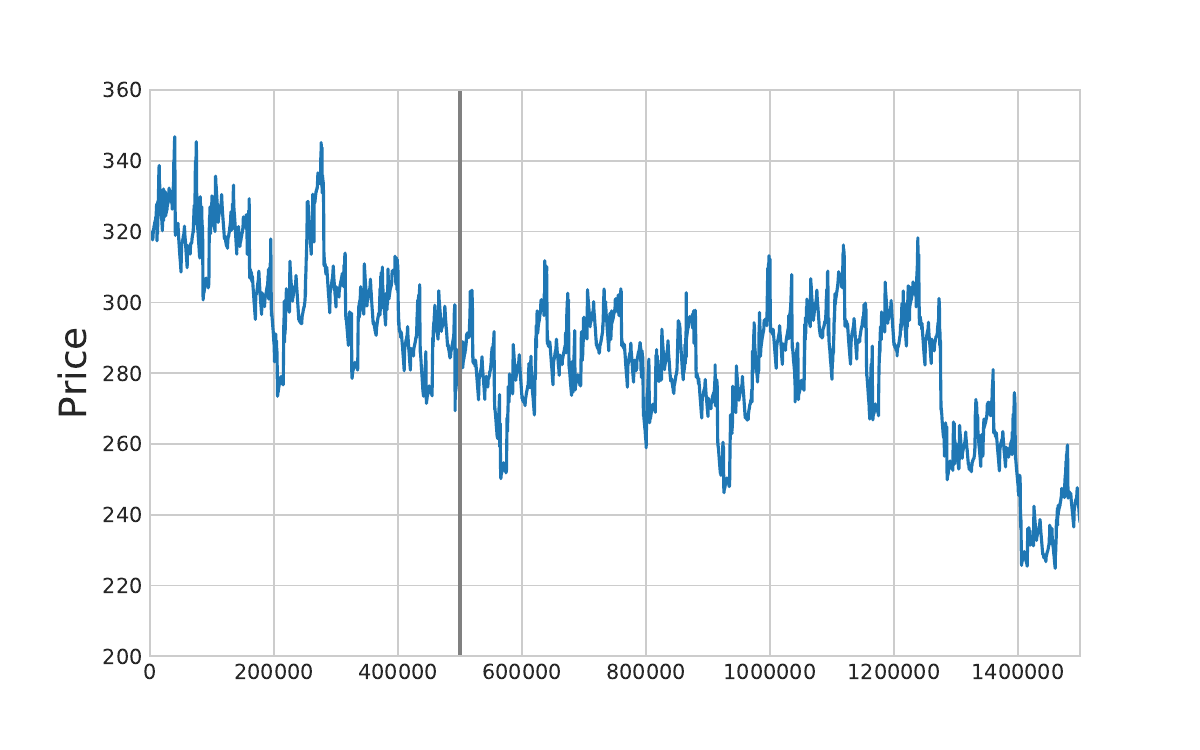}}
    \subcaptionbox[Stream 8]{\label{fig:appendix:synthetic:h}}
    [0.24\textwidth]{\includegraphics[clip, trim=0cm 0.5cm 0.5cm 1.3cm, width=0.99\linewidth]{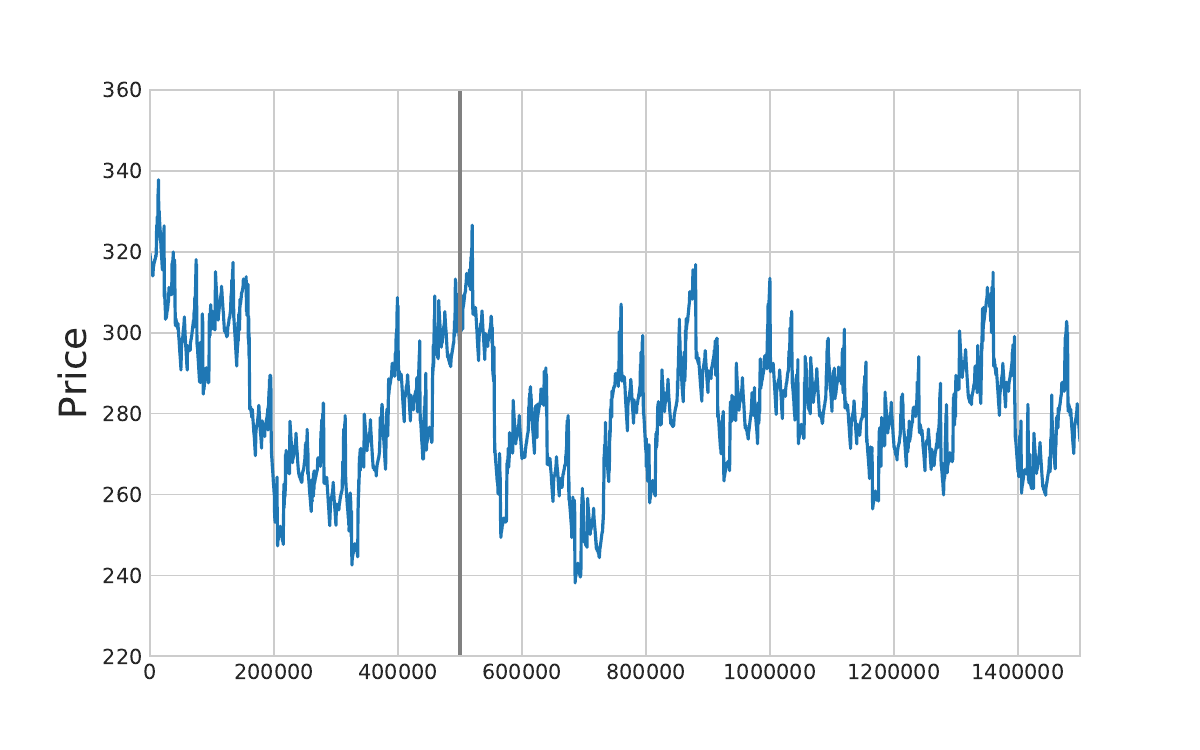}}
    \caption[Reconstructed close price from synthetic sets 1-8.]{Reconstructed close price series for the first eight generated synthetic data streams.}
    \label{fig:appendix:synthetic:all}
\end{figure*}

\begin{figure*}[hbt!]
    \centering
    \subcaptionbox[Stream 9]{\label{fig:appendix:synthetic:i}}
    [0.24\textwidth]{\includegraphics[clip, trim=0cm 0.5cm 0.5cm 1.3cm, width=0.99\linewidth]{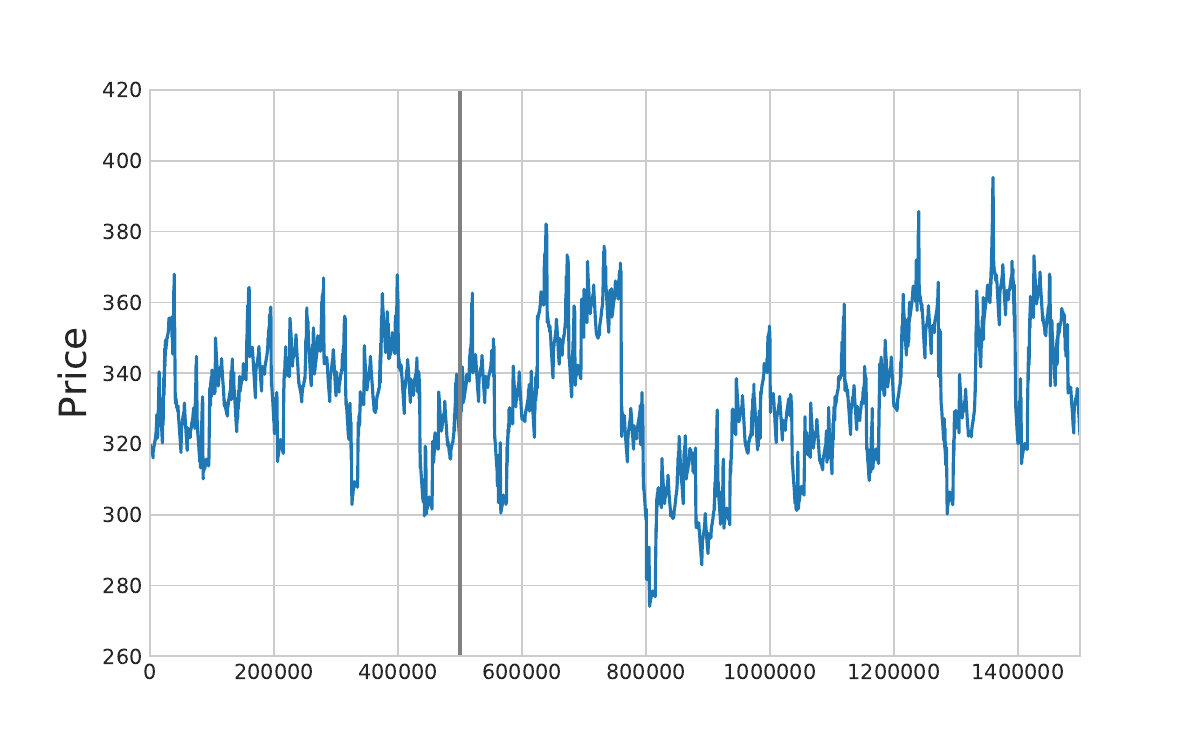}}
    \subcaptionbox[Stream 10]{\label{fig:appendix:synthetic:j}}
    [0.24\textwidth]{\includegraphics[clip, trim=0cm 0.5cm 0.5cm 1.3cm, width=0.99\linewidth]{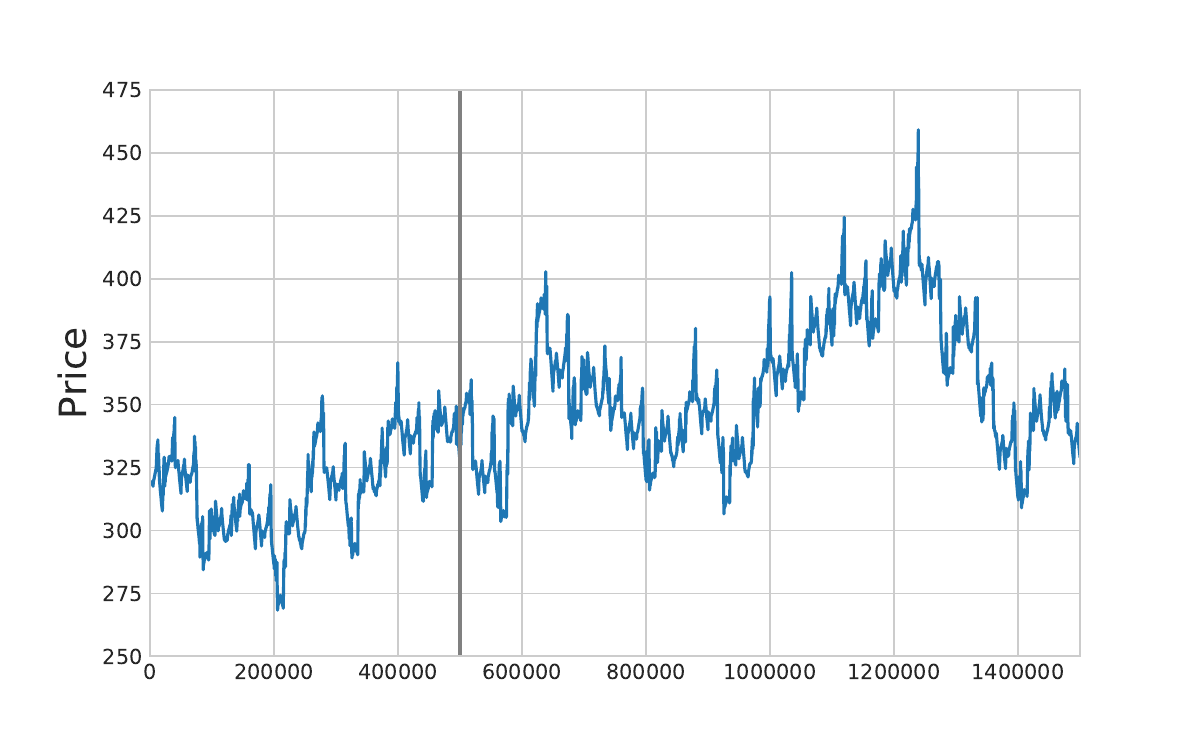}}
    \subcaptionbox[Stream 11]{\label{fig:appendix:synthetic:k}}
    [0.24\textwidth]{\includegraphics[clip, trim=0cm 0.5cm 0.5cm 1.3cm, width=0.99\linewidth]{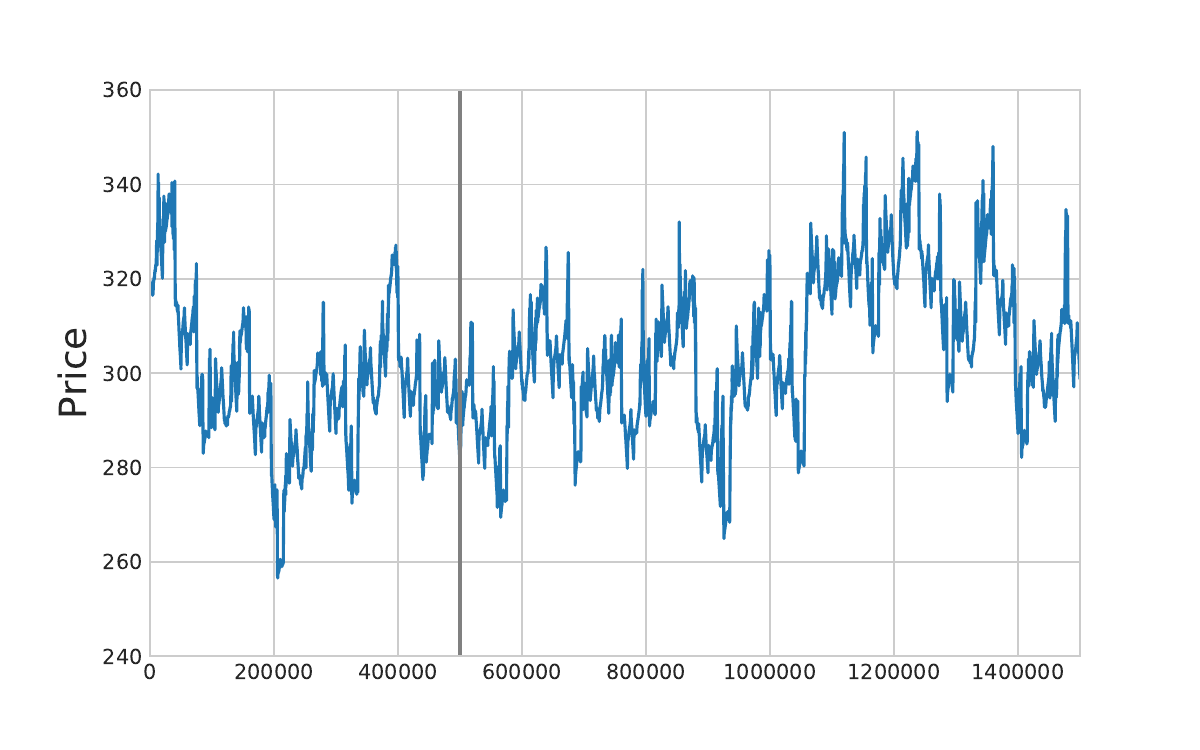}}
    \subcaptionbox[Stream 12]{\label{fig:appendix:synthetic:l}}
    [0.24\textwidth]{\includegraphics[clip, trim=0cm 0.5cm 0.5cm 1.3cm, width=0.99\linewidth]{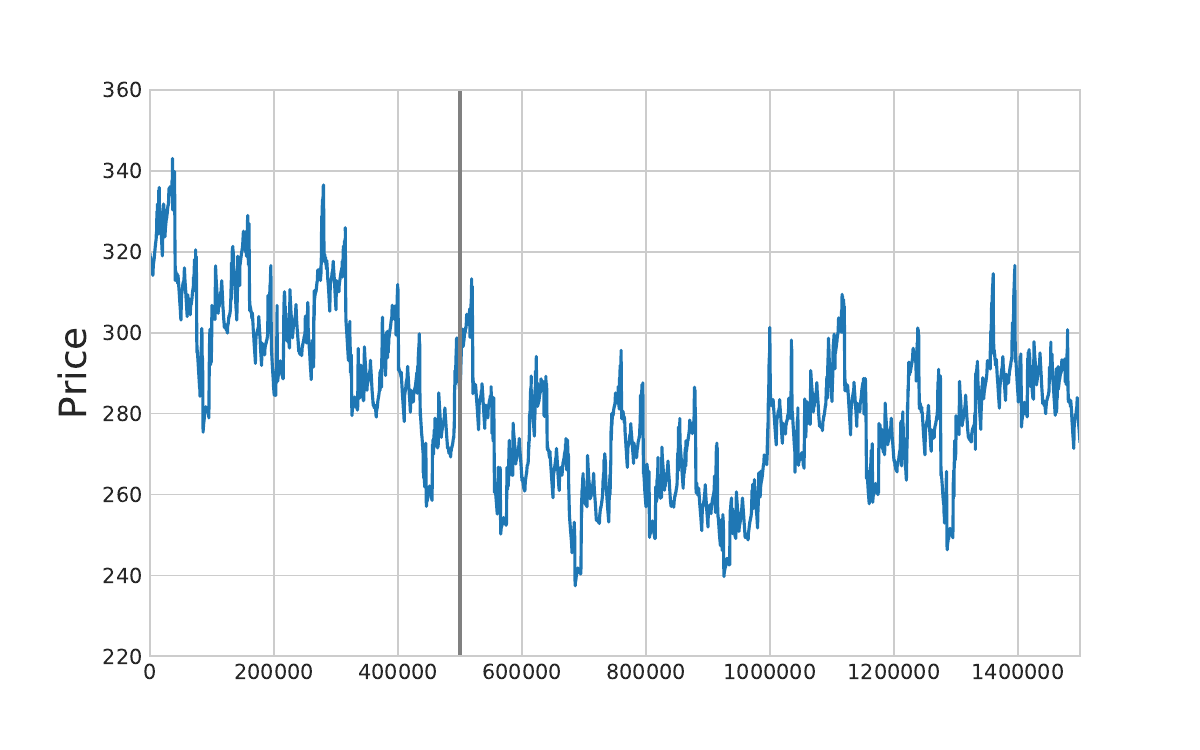}}
    \subcaptionbox[Stream 13]{\label{fig:appendix:synthetic:m}}
    [0.24\textwidth]{\includegraphics[clip, trim=0cm 0.5cm 0.5cm 1.3cm, width=0.99\linewidth]{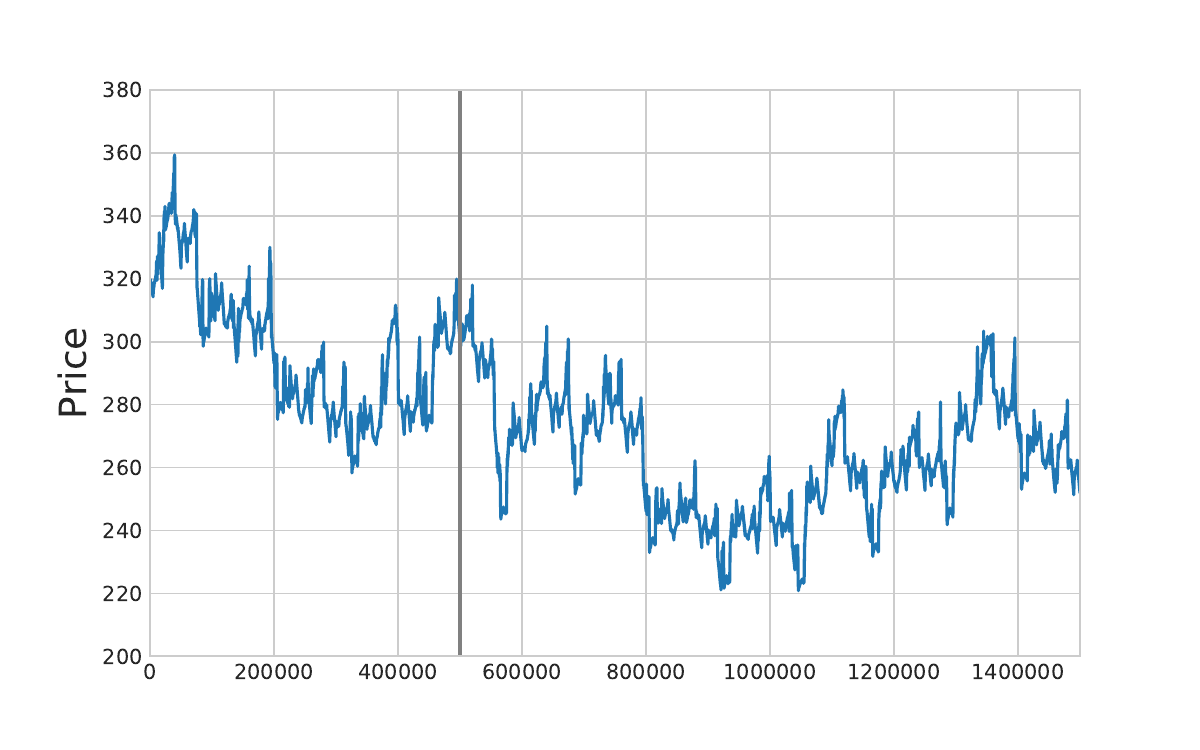}}
    \subcaptionbox[Stream 14]{\label{fig:appendix:synthetic:n}}
    [0.24\textwidth]{\includegraphics[clip, trim=0cm 0.5cm 0.5cm 1.3cm, width=0.99\linewidth]{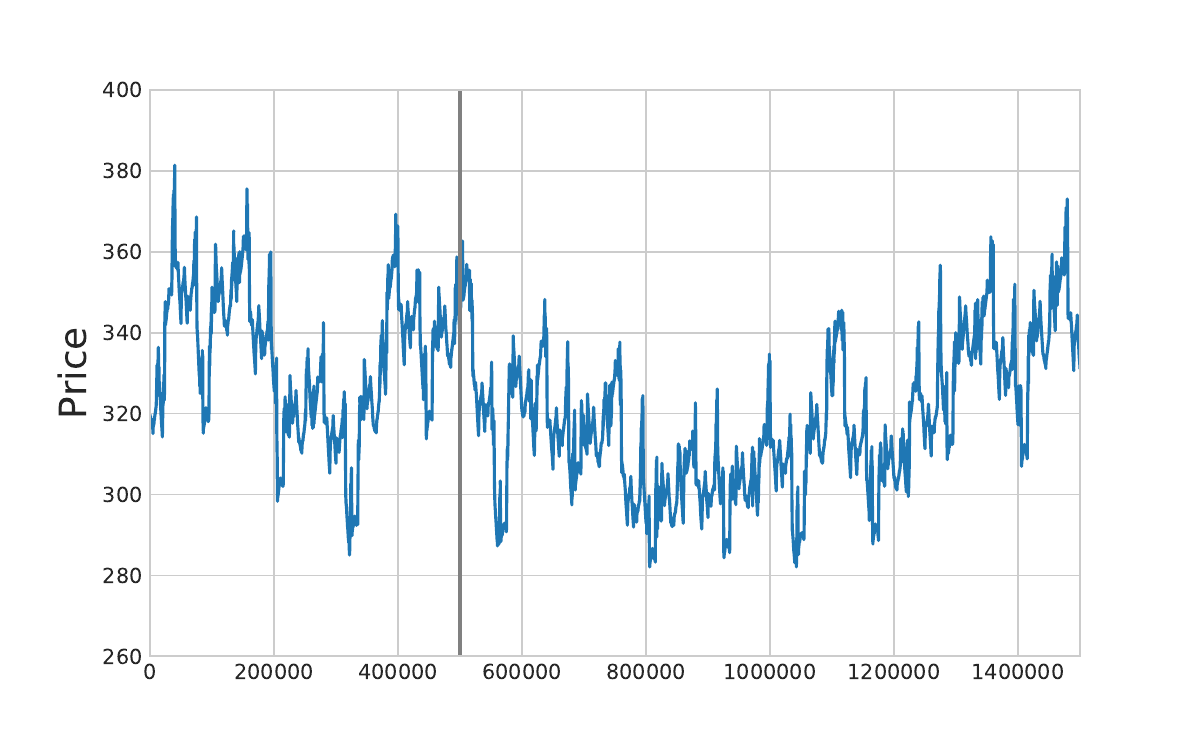}}
    \subcaptionbox[Stream 15]{\label{fig:appendix:synthetic:o}}
    [0.24\textwidth]{\includegraphics[clip, trim=0cm 0.5cm 0.5cm 1.3cm, width=0.99\linewidth]{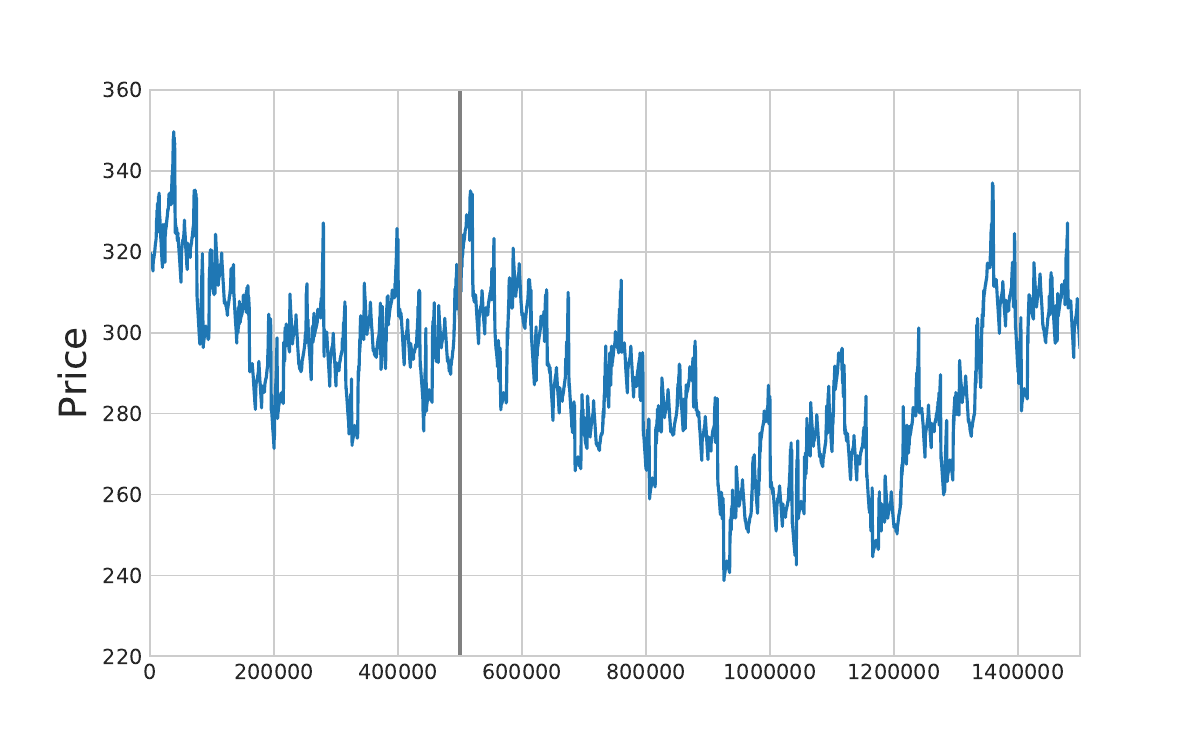}}
    \subcaptionbox[Stream 16]{\label{fig:appendix:synthetic:p}}
    [0.24\textwidth]{\includegraphics[clip, trim=0cm 0.5cm 0.5cm 1.3cm, width=0.99\linewidth]{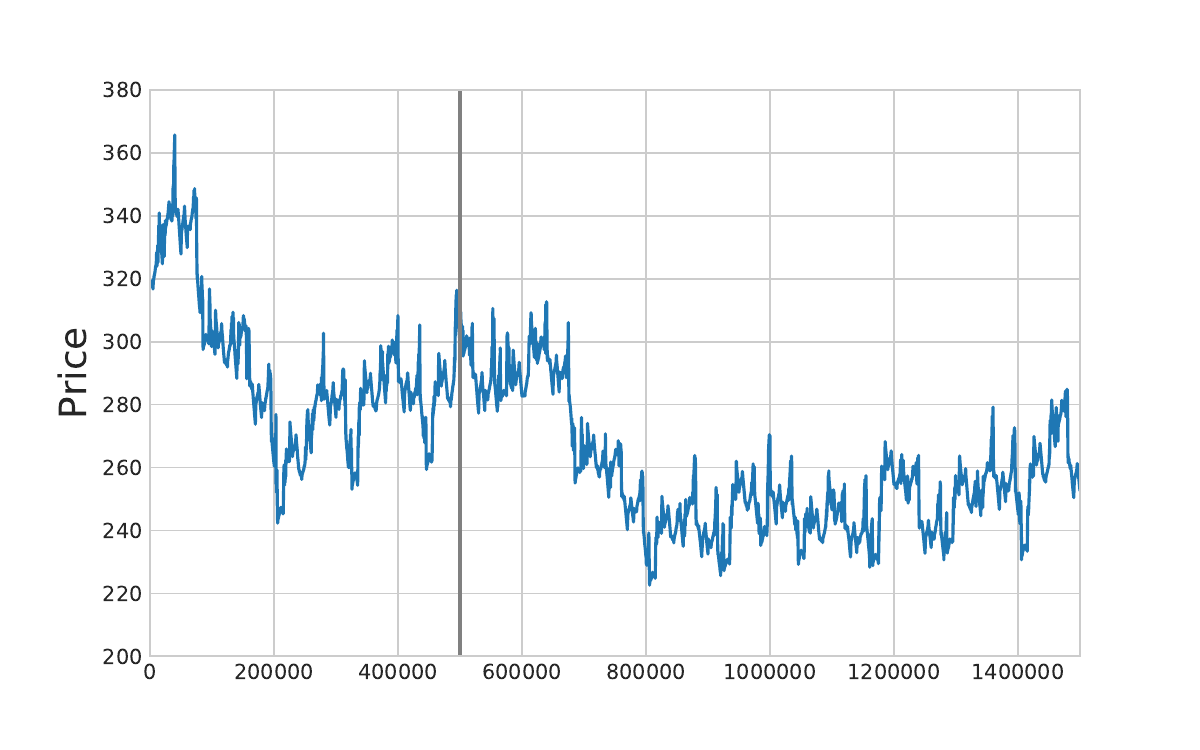}}
    \caption[Reconstructed close price from synthetic sets 9-16.]{Reconstructed close price series for synthetic data streams 9 through 16.}
    \label{fig:appendix:synthetic:all2}
\end{figure*}

\begin{figure*}[hbt!]
    \centering
    \subcaptionbox[Stream 17]{\label{fig:appendix:synthetic:o2}}
    [0.24\textwidth]{\includegraphics[clip, trim=0cm 0.5cm 0.5cm 1.3cm, width=0.99\linewidth]{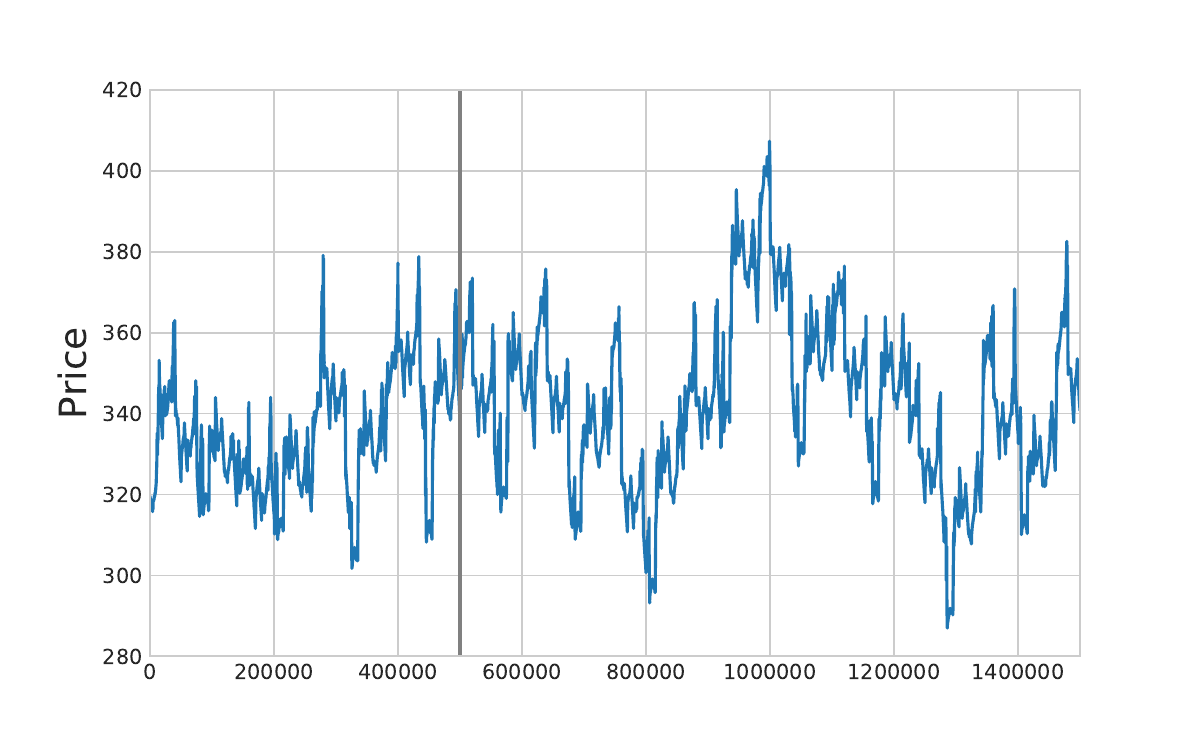}}
    \subcaptionbox[Stream 18]{\label{fig:appendix:synthetic:q}}
    [0.24\textwidth]{\includegraphics[clip, trim=0cm 0.5cm 0.5cm 1.3cm, width=0.99\linewidth]{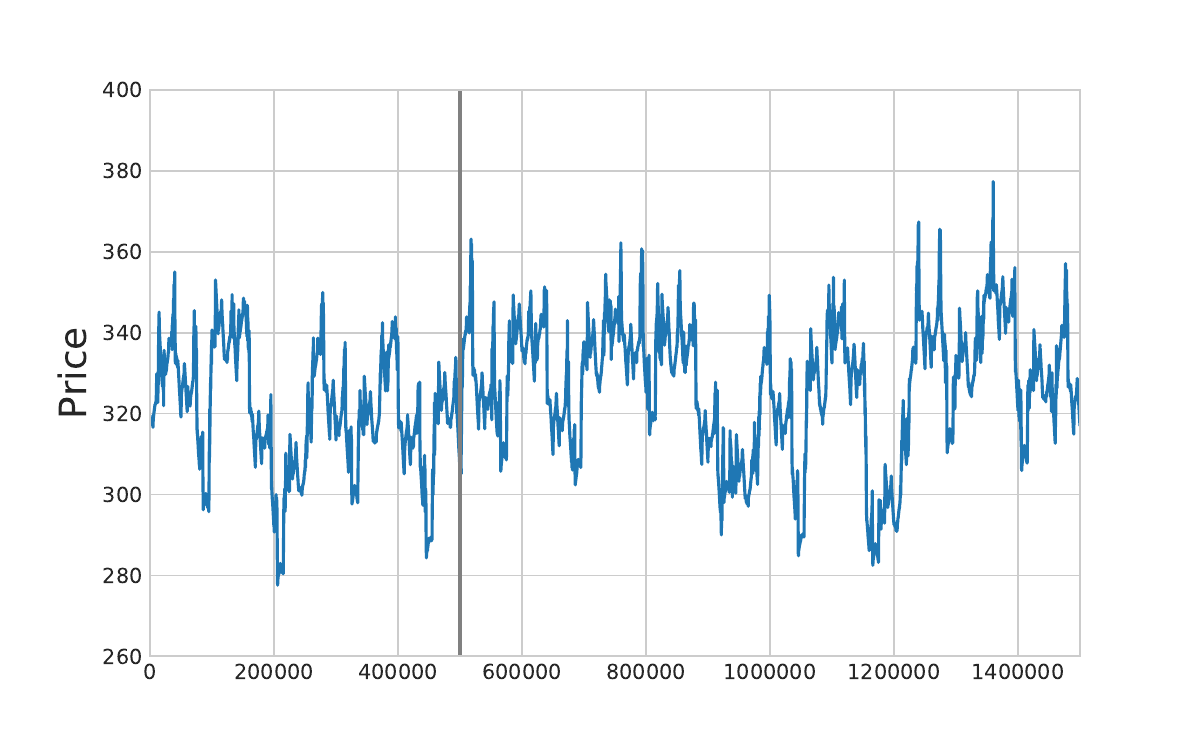}}
    \subcaptionbox[Stream 19]{\label{fig:appendix:synthetic:r}}
    [0.24\textwidth]{\includegraphics[clip, trim=0cm 0.5cm 0.5cm 1.3cm, width=0.99\linewidth]{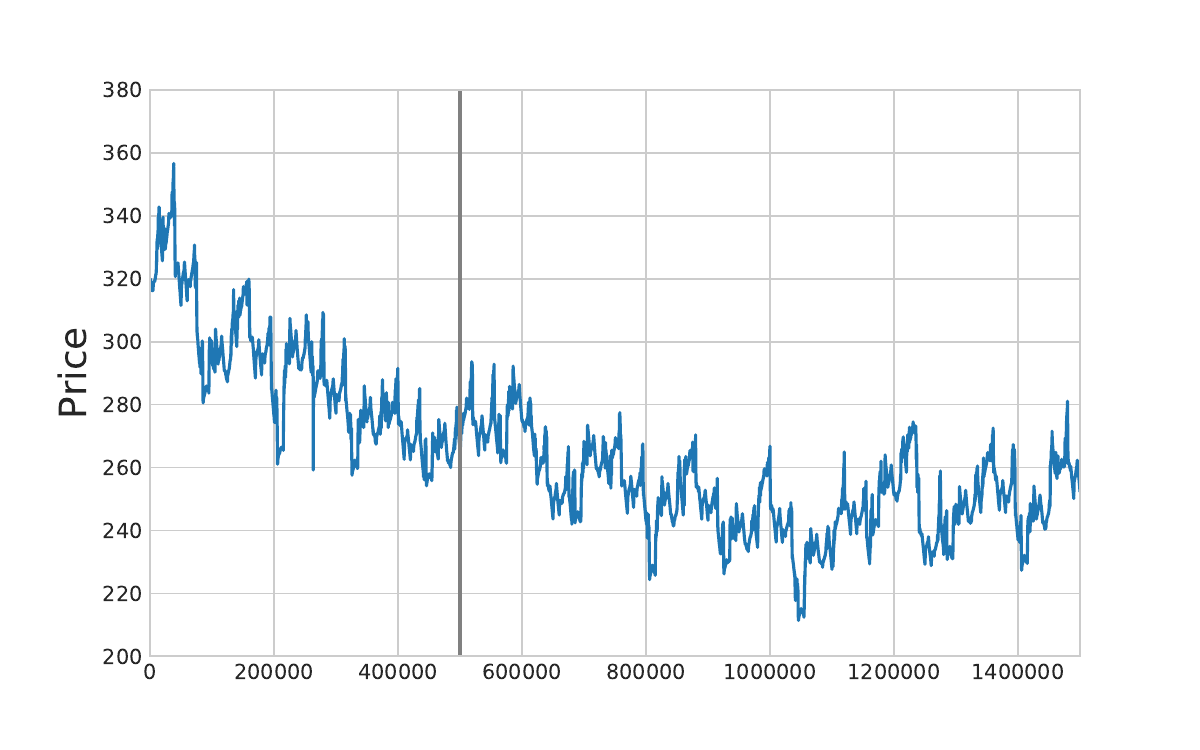}}
    \subcaptionbox[Stream 20]{\label{fig:appendix:synthetic:s}}
    [0.24\textwidth]{\includegraphics[clip, trim=0cm 0.5cm 0.5cm 1.3cm, width=0.99\linewidth]{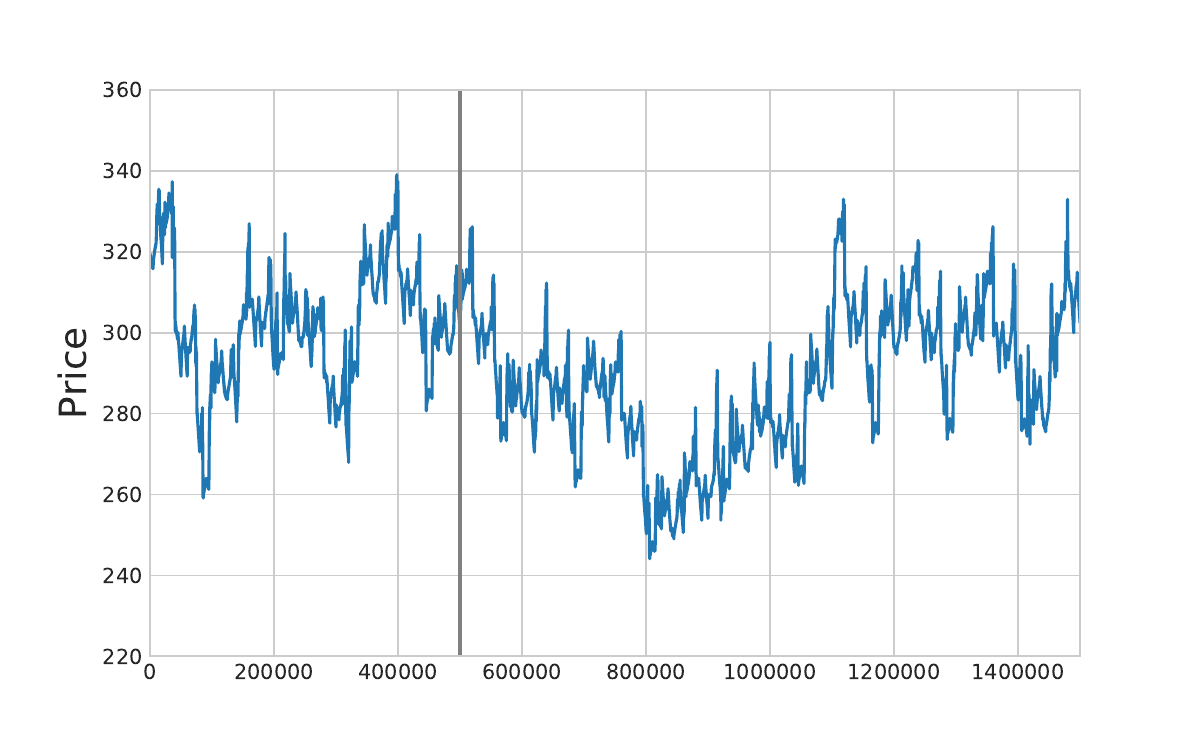}}
    \subcaptionbox[Stream 21]{\label{fig:appendix:synthetic:t}}
    [0.24\textwidth]{\includegraphics[clip, trim=0cm 0.5cm 0.5cm 1.3cm, width=0.99\linewidth]{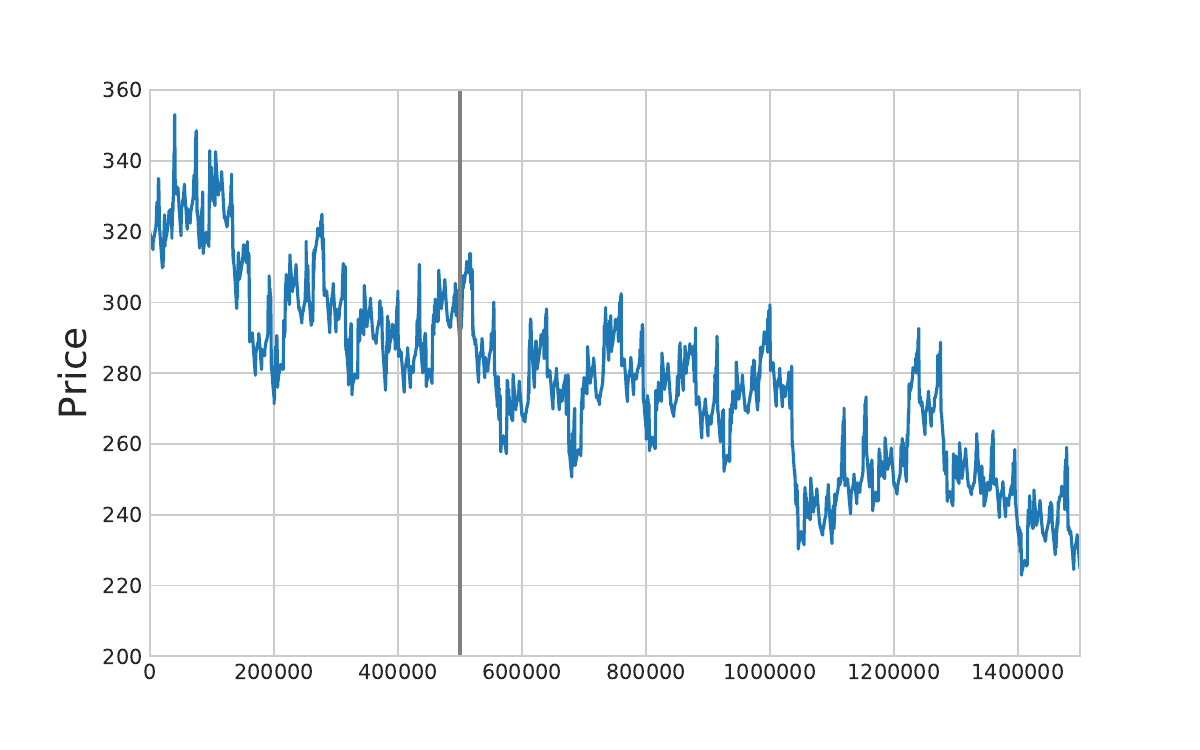}}
    \subcaptionbox[Stream 22]{\label{fig:appendix:synthetic:v}}
    [0.24\textwidth]{\includegraphics[clip, trim=0cm 0.5cm 0.5cm 1.3cm, width=0.99\linewidth]{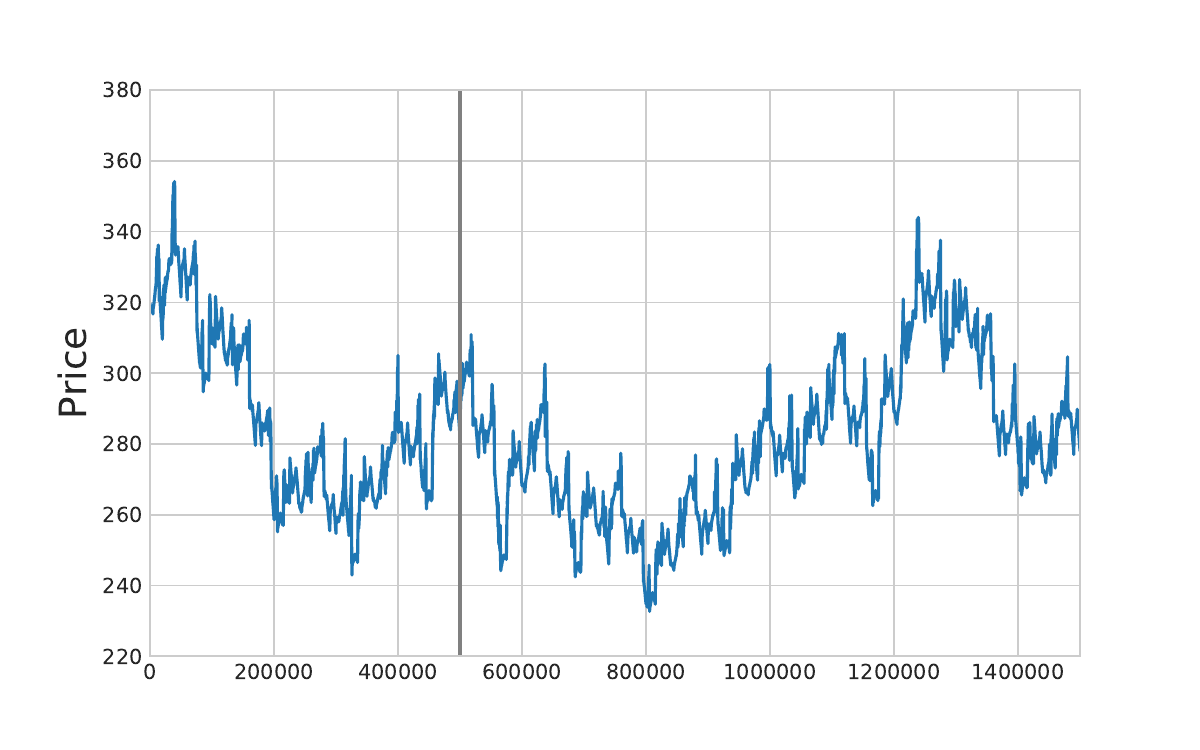}}
    \subcaptionbox[Stream 23]{\label{fig:appendix:synthetic:w}}
    [0.24\textwidth]{\includegraphics[clip, trim=0cm 0.5cm 0.5cm 1.3cm, width=0.99\linewidth]{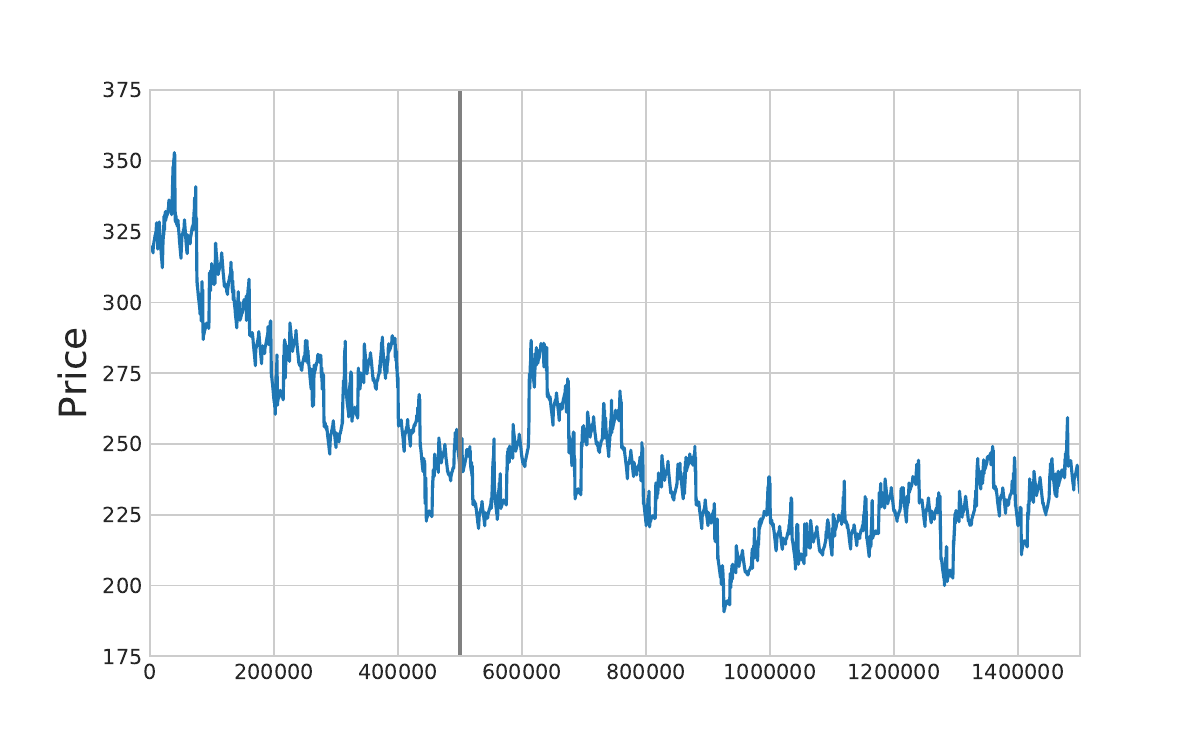}}
    \subcaptionbox[Stream 24]{\label{fig:appendix:synthetic:x}}
    [0.24\textwidth]{\includegraphics[clip, trim=0cm 0.5cm 0.5cm 1.3cm, width=0.99\linewidth]{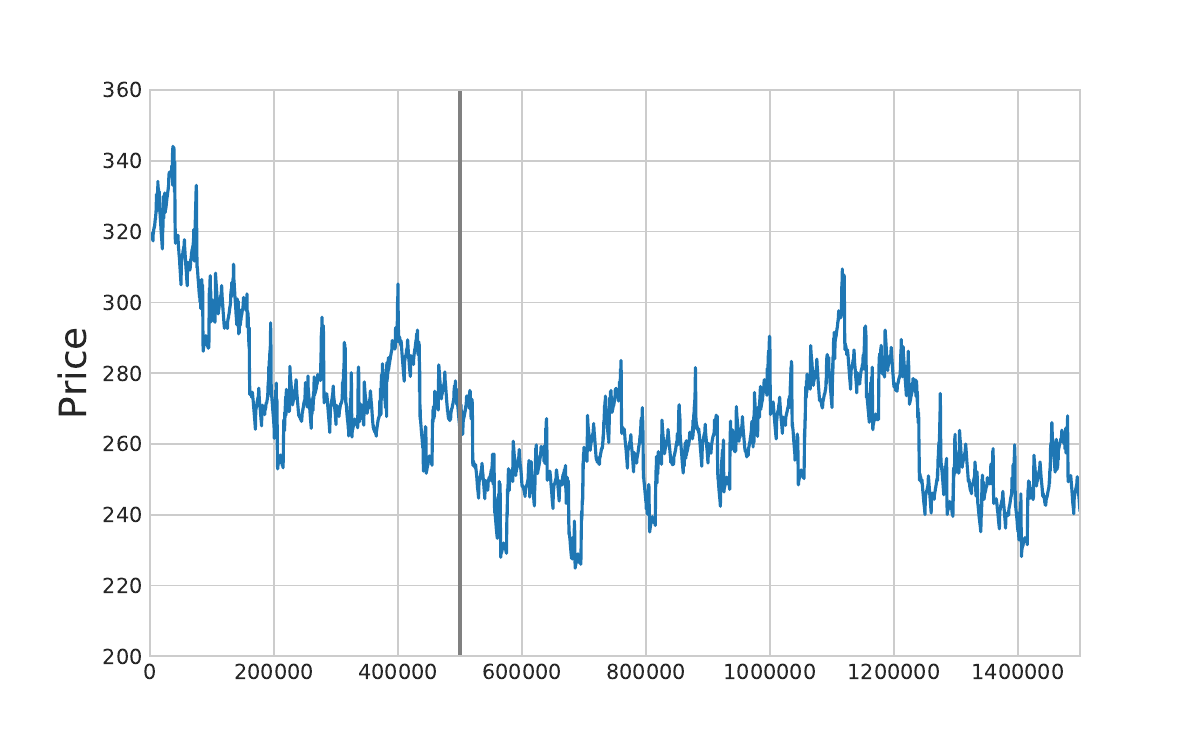}}
    \caption[Reconstructed close price from synthetic sets 17-24.]{Reconstructed close price series for synthetic data streams 17 through 24.}
    \label{fig:appendix:synthetic:all3}
\end{figure*}

\begin{figure*}[hbt!]
    \centering
    \subcaptionbox[Stream 25]{\label{fig:appendix:synthetic:y}}
    [0.24\textwidth]{\includegraphics[clip, trim=0cm 0.5cm 0.5cm 1.3cm, width=0.99\linewidth]{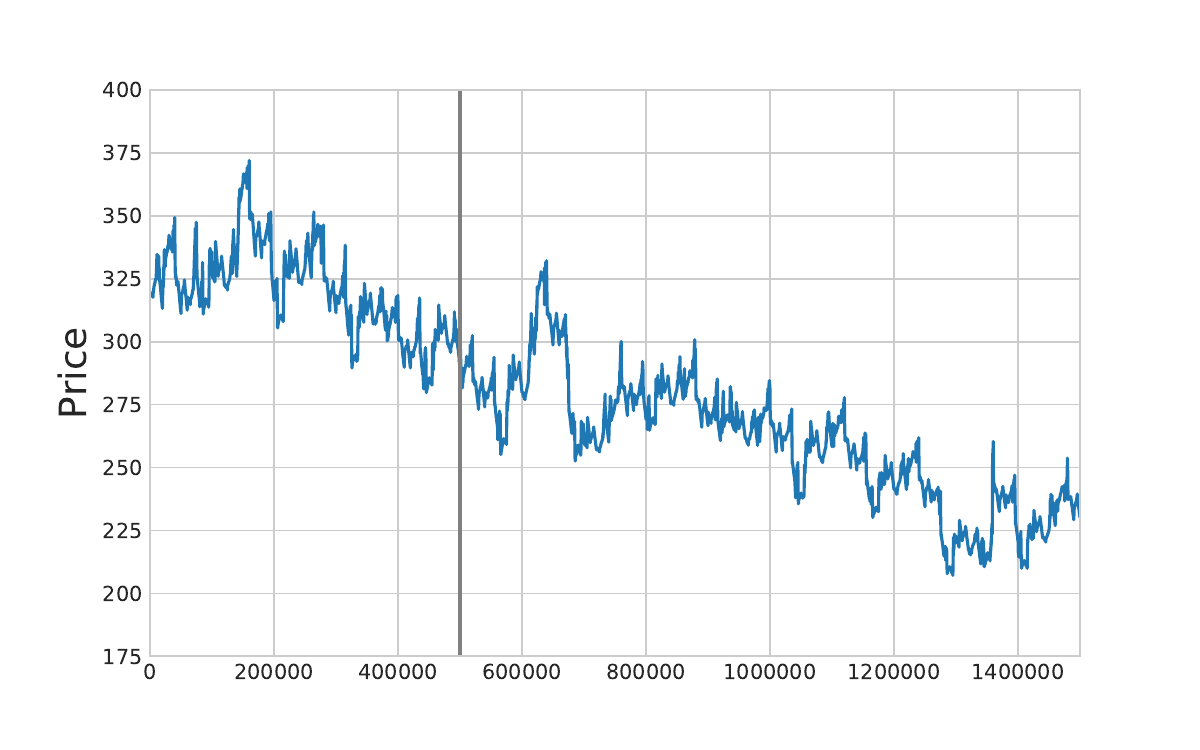}}
    \subcaptionbox[Stream 26]{\label{fig:appendix:synthetic:z}}
    [0.24\textwidth]{\includegraphics[clip, trim=0cm 0.5cm 0.5cm 1.3cm, width=0.99\linewidth]{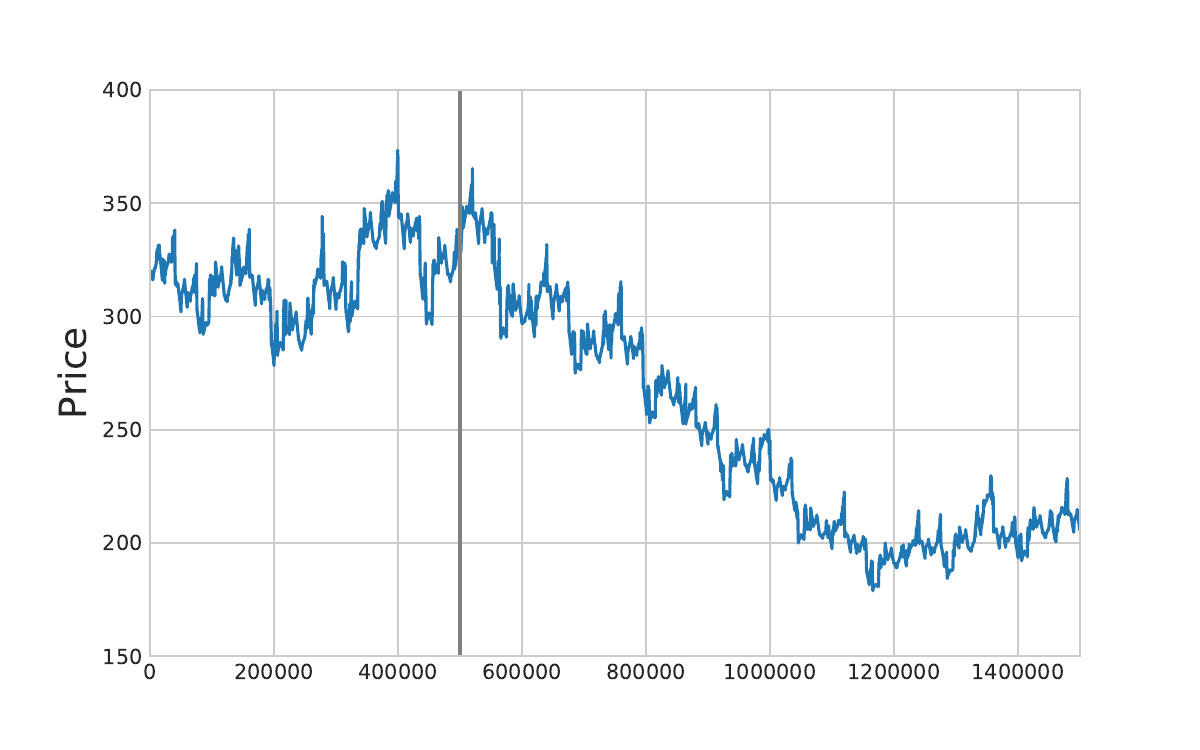}}
    \subcaptionbox[Stream 27]{\label{fig:appendix:synthetic:z2}}
    [0.24\textwidth]{\includegraphics[clip, trim=0cm 0.5cm 0.5cm 1.3cm, width=0.99\linewidth]{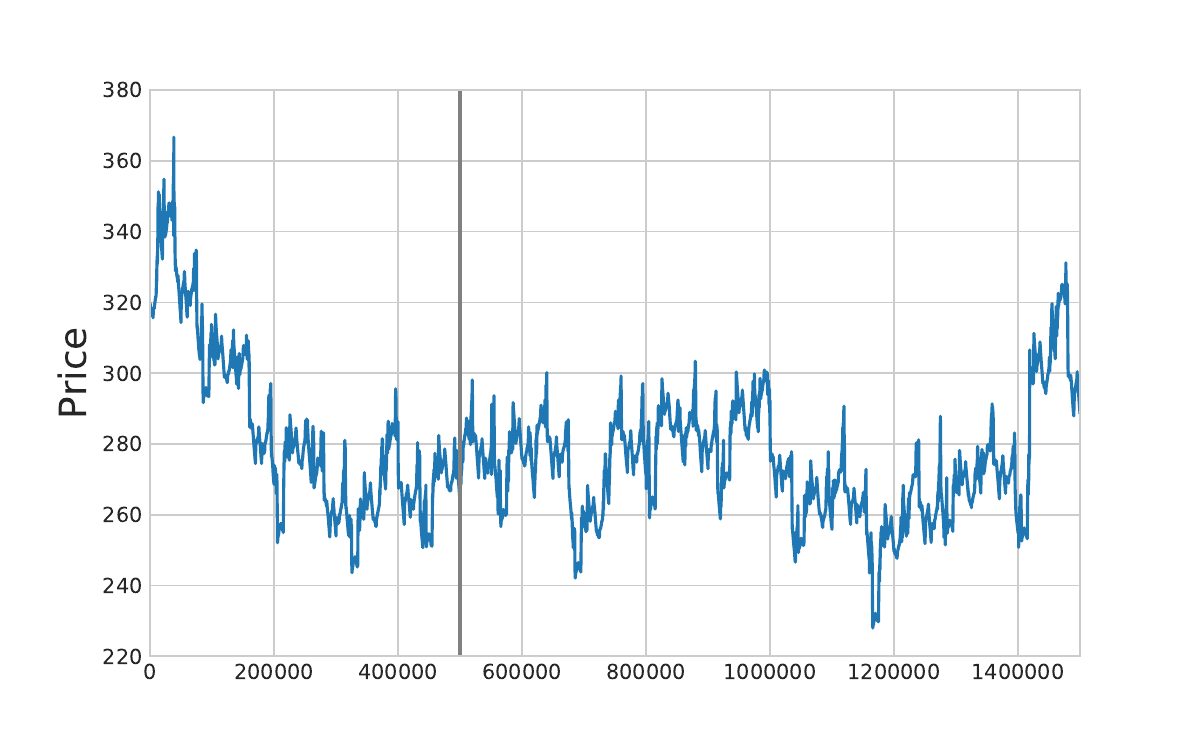}}
    \subcaptionbox[Stream 28]{\label{fig:appendix:synthetic:z3}}
    [0.24\textwidth]{\includegraphics[clip, trim=0cm 0.5cm 0.5cm 1.3cm, width=0.99\linewidth]{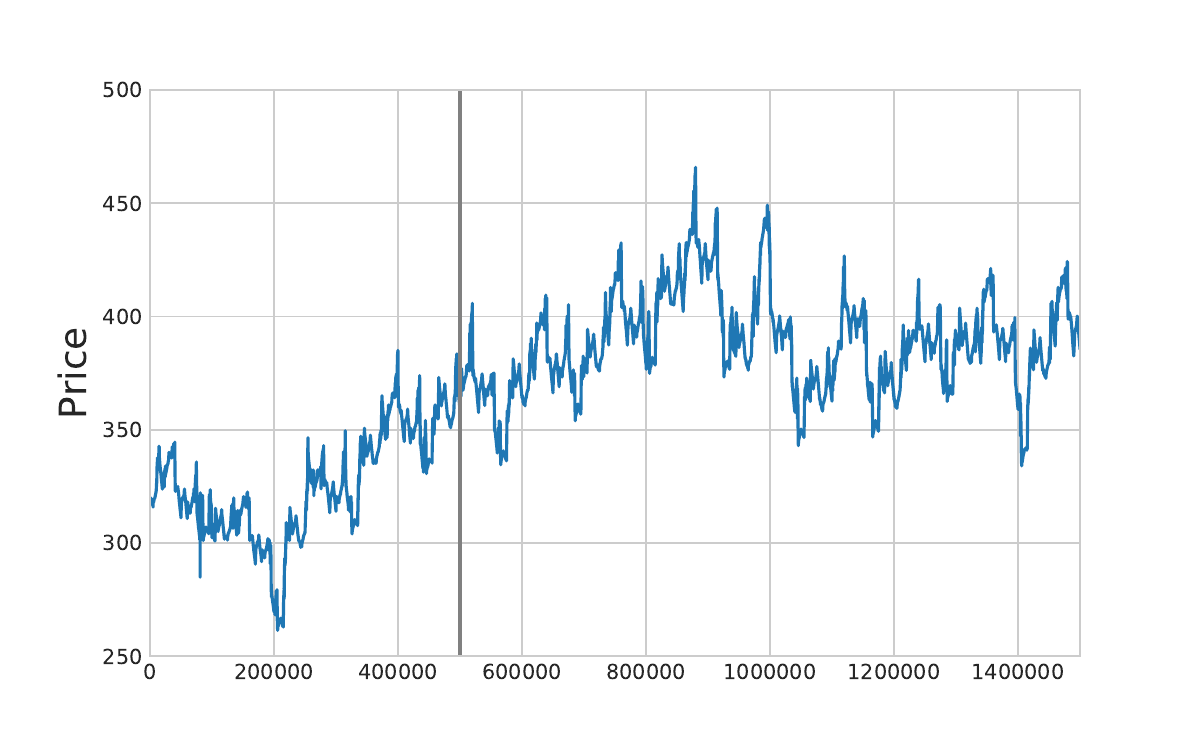}}
    \subcaptionbox[Stream 29]{\label{fig:appendix:synthetic:z4}}
    [0.24\textwidth]{\includegraphics[clip, trim=0cm 0.5cm 0.5cm 1.3cm, width=0.99\linewidth]{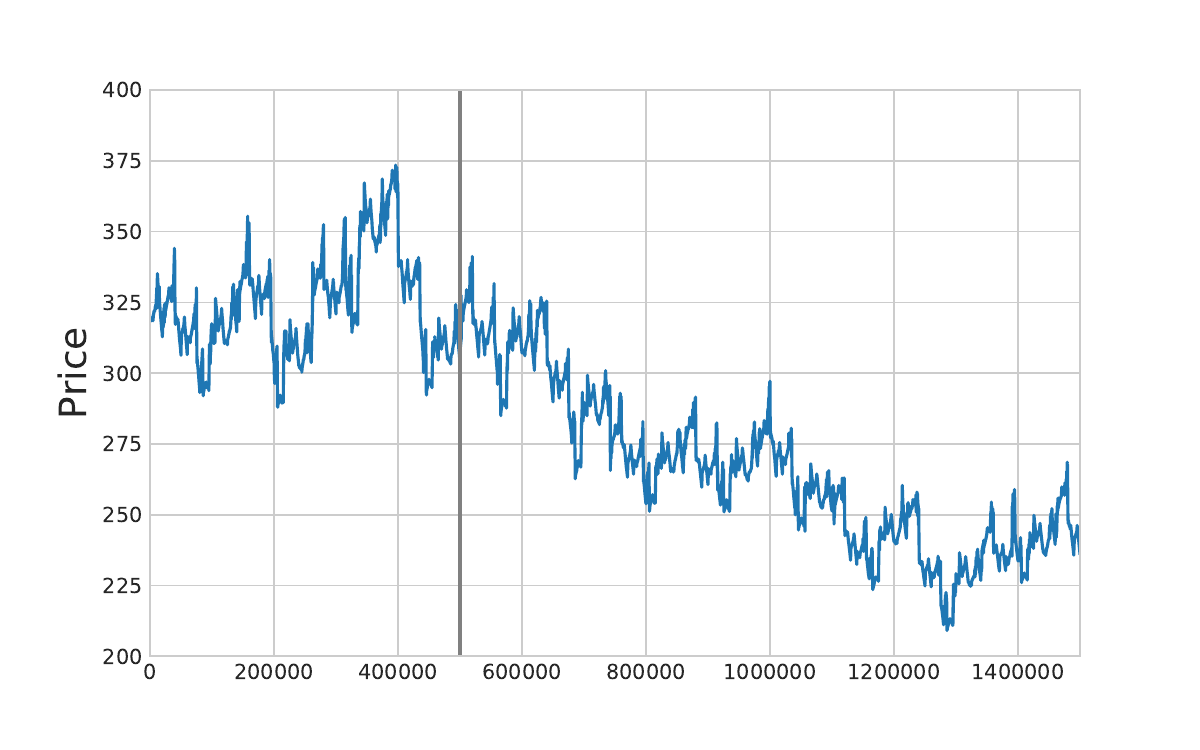}}
    \subcaptionbox[Stream 30]{\label{fig:appendix:synthetic:z5}}
    [0.24\textwidth]{\includegraphics[clip, trim=0cm 0.5cm 0.5cm 1.3cm, width=0.99\linewidth]{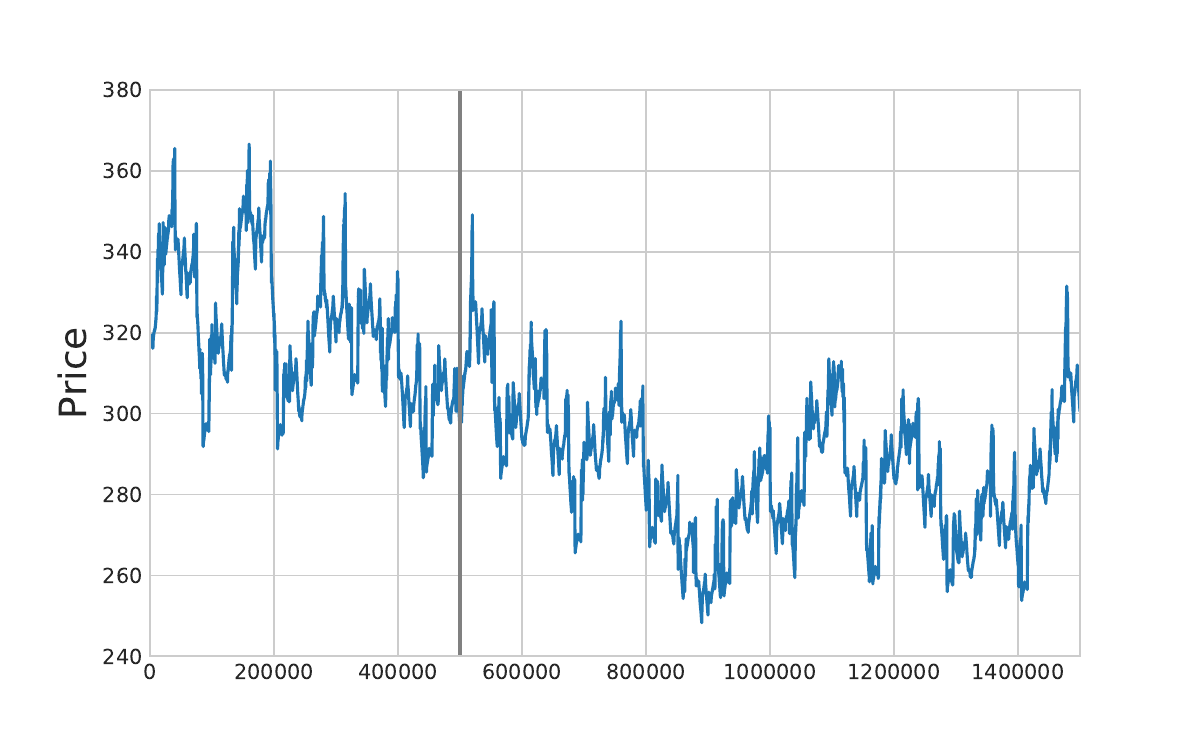}}
    \caption[Reconstructed close price from synthetic sets 25-30.]{Reconstructed close price series for synthetic data streams 25 through 30.}
    \label{fig:appendix:synthetic:all4}
\end{figure*}

Following the setup described in Section~\ref{sec:expsetup}, we used ProteuS to generate 30 unique semi-synthetic data streams, each containing 1.5 million data points. The final reconstructed close prices for all 30 streams are shown in Figures~\ref{fig:appendix:synthetic:all}-\ref{fig:appendix:synthetic:all4}.

Each stream consists of a series of price returns simulated from the ARMA-GARCH models, which were then reconstructed into price series. Each one adheres to the ground truth transition map specified in Section~\ref{sec:expsetup}, with regime changes occurring at pre-defined intervals.

\subsection{Analysis of the Feature Space}
\label{sec:analysis_feature_space}

The final stage of ProteuS is feature engineering from the reconstructed OHLC price series. % as described in Section 3. 
We computed 18 features using different technical indicators, selected for their widespread use in financial forecasting and their ability to capture diverse aspects of market dynamics, including trend, momentum, and volatility.
Table~\ref{table:stats_syntheticdataset} provides descriptive statistics for this feature set, calculated over the first one million instances of a representative synthetic stream. 

\begin{table}[htbp]
  \resizebox{.95\textwidth}{!}{
  \centering
  \begin{minipage}{\textwidth}
        \begin{tabular}{lrrrrrrr}
            \toprule
            {} &           Mean &           Std. &           Min. &            25\% &           50\% &            75\% &           Max. \\
            \midrule
            RSI$_{10}$           &   50.620 &   13.179 &    1.801 &   41.957 &   49.474 &   57.940 &   99.999 \\
            WILLR$_{10}$         &  -47.700 &   35.520 & -100.000 &  -79.286 &  -48.068 &  -12.532 &    0.000 \\
            MACD        &   -0.000 &    0.070 &   -2.718 &   -0.016 &   -0.002 &    0.013 &    1.953 \\
            CCI$_{10}$          &    0.794 &  101.982 & -333.333 &  -78.932 &    0.878 &   86.799 &  333.333 \\
            MOM$_{10}$          &   -0.001 &    0.172 &   -7.050 &   -0.034 &   -0.001 &    0.032 &    7.494 \\
            SK      &   53.453 &   29.211 &    0.000 &   29.119 &   53.172 &   78.550 &  100.000 \\
            SD      &   53.453 &   26.516 &    0.000 &   32.360 &   52.857 &   74.788 &  100.000 \\
            SMA$_{5}$           &  260.808 &   21.003 &  220.642 &  242.941 &  260.208 &  274.216 &  333.426 \\
            SMA$_{10}$          &  260.809 &   21.003 &  220.677 &  242.940 &  260.208 &  274.217 &  333.404 \\
            WMA$_{10}$          &  260.808 &   21.003 &  220.657 &  242.942 &  260.209 &  274.215 &  333.428 \\
            EMA$_{10}$          &  260.809 &   21.003 &  220.668 &  242.940 &  260.208 &  274.215 &  333.419 \\
            TRIMA$_{10}$        &  260.809 &   21.003 &  220.684 &  242.940 &  260.208 &  274.217 &  333.404 \\
            ADX$_{10}$          &   32.167 &   17.341 &    2.630 &   19.003 &   28.266 &   41.481 &  100.000 \\
            Bollinger$_{upperband}$ &  260.937 &   21.030 &  221.003 &  243.096 &  260.310 &  274.336 &  333.716 \\
            Bollinger$_{lowerband}$ &  260.680 &   20.978 &  220.295 &  242.811 &  260.113 &  274.083 &  333.117 \\
            ROC$_{10}$          &   -0.000 &    0.065 &   -2.601 &   -0.013 &   -0.000 &    0.012 &    2.855 \\
            Aroon$_{DOWN}$  &   51.430 &   37.307 &    0.000 &   10.000 &   50.000 &   90.000 &  100.000 \\
            Aroon$_{UP}$   &   53.729 &   37.684 &    0.000 &   20.000 &   50.000 &   90.000 &  100.000 \\
            \bottomrule
        \end{tabular}
      \end{minipage}}
 \caption[Descriptive statistics of 1 million instances in the first synthetic stream.]{Descriptive statistics of the first 1 million instances of the first synthetic stream created. \label{table:stats_syntheticdataset}}
\end{table}

A key observation from these statistics is the distinction between scaled indicators (e.g., RSI, WILLR, which are bounded between 0 and 100 or -100 and 0) and unscaled indicators (e.g., SMA, MACD, MOM), which are dependent on the price level and can exhibit significant variance. 
The binary target label is well-balanced across the streams, with the proportion of the downtrend class (label 0) typically falling between 45\% and 55\%.

\subsection{Discussion on State Separability}
\label{sec:discussion_separability}

While ProteuS generates data from four distinct ground truth states, the inherent complexity of the simulation process presents several challenges. A primary issue is that the generative models, while fitted on specific historical data, are used to generate a continuous time series where the output of one model becomes the input for the next during a transition. This process means that the models are exposed to data distributions for which they were not explicitly trained, potentially leading to unforeseen behaviors in the generated stream.

One significant challenge we encountered was the tendency for the reconstructed price series to exhibit explosive, unrealistic trends over the long-term simulation of 1.5 million data points. To mitigate this, we carefully selected the initial ETFs to have near-zero mean returns, ensuring that the generated series remained relatively stable and that the technical indicators behaved realistically.
 
\begin{figure}[ht] 
    \centering
    \includegraphics[trim={1.1cm 0cm 1.6cm 0.5cm},clip,
    width=0.99\linewidth]{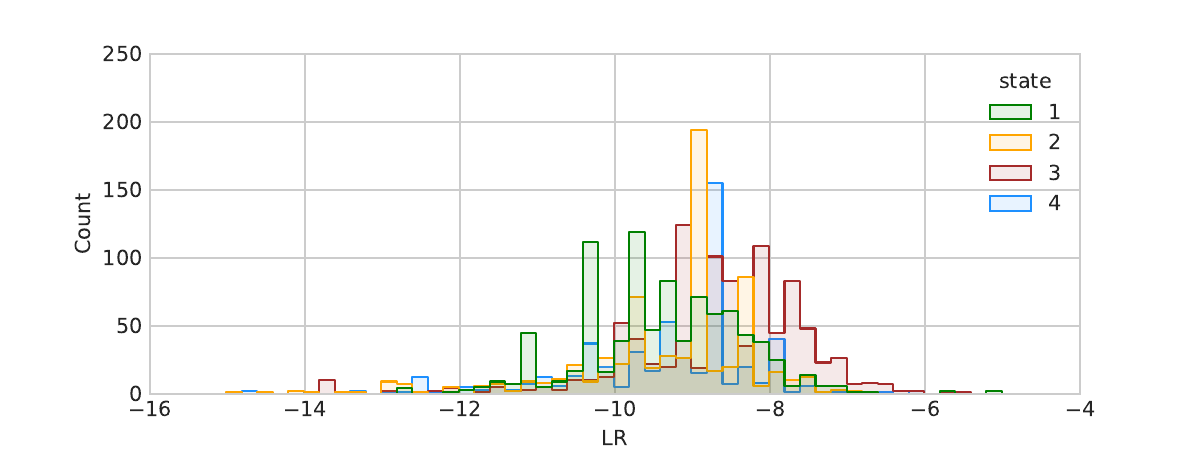}
    \caption[Histogram of returns over the four states used.]{The distribution of logarithmic returns for the four raw market states, illustrating the inherent similarities and differences in their statistical properties.}
    \label{fig:histogram1}
\end{figure}

Furthermore, the separability of the market states in the final feature space is not guaranteed. To assess the inherent difficulty of the classification task, we conducted a validation analysis using the visualization framework proposed by \cite{Tsang2020}. Figure~\ref{fig:histogram1} shows the distribution of logarithmic returns for the four raw market states, revealing considerable overlap even before the simulation process. 

\begin{figure*}[hbt!]
\centering
    \subcaptionbox[Short Subcaption A]{Raw states \label{fig:cluster1}}
    [0.495\textwidth]{
    \includegraphics[clip, width=0.99\linewidth]{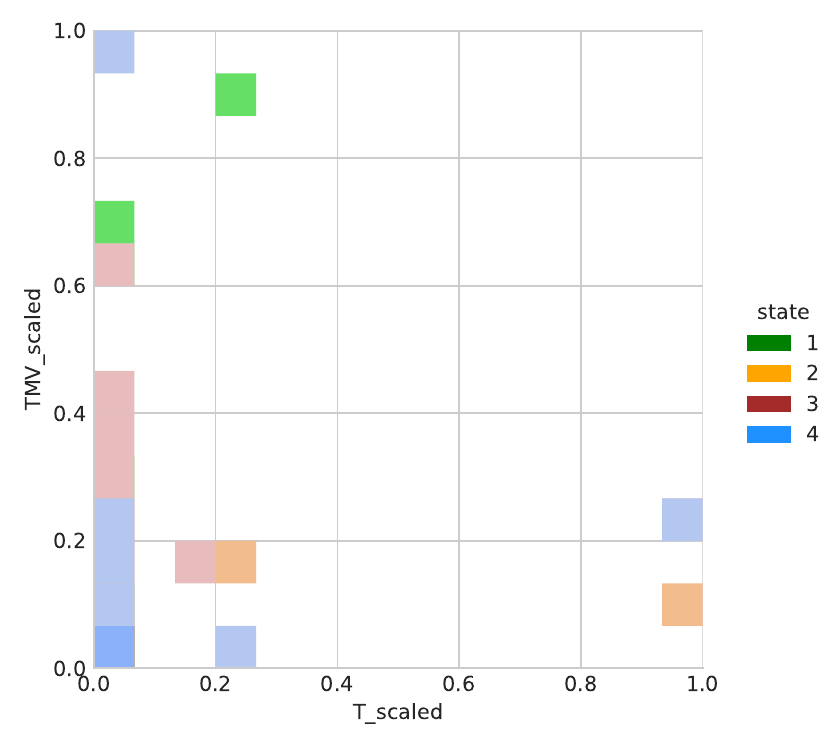}}
    \centering
    \subcaptionbox[Short Subcaption B]{Synthetic states \label{fig:cluster2}}
    [0.495\textwidth]{
    \includegraphics[clip, width=0.99\linewidth]{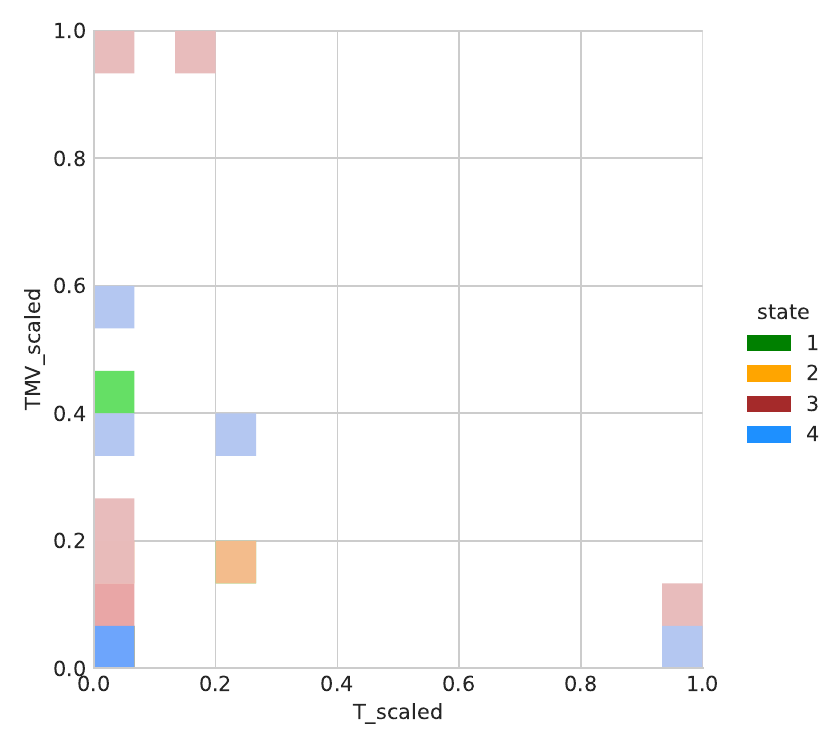}}
  \caption[Representation of the fours states used for the synthetic sets.]{A comparison of the separability of the four market states in the raw historical data versus the generated synthetic data, using the visualization framework from \cite{Tsang2020}.}
  \label{fig:2Dhistograms}
\end{figure*}

This challenge is amplified in the final synthetic data, as shown in Figure~\ref{fig:2Dhistograms}, where the synthetic states (Figure~\ref{fig:cluster2}) exhibit a much greater degree of overlap than their raw counterparts (Figure~\ref{fig:cluster1}).

\begin{figure}[hbt] 
    \centering
    \includegraphics[trim={1.5cm 0cm 1.5cm 0cm},clip, width=0.8\linewidth]{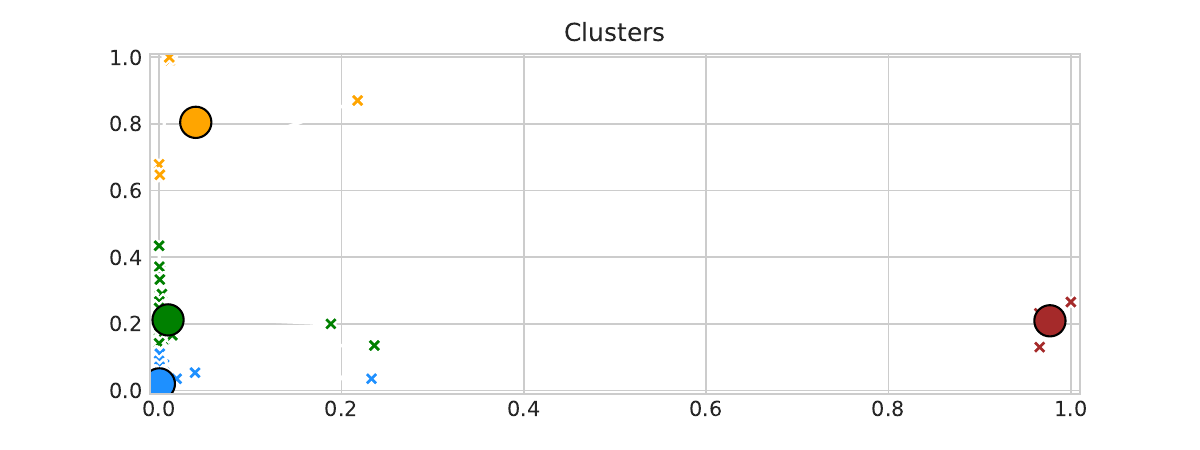}
    \caption[K-means cluster centers for k=4.]{The centroids found by a k-means algorithm (k=4) on the synthetic data from Figure~\ref{fig:cluster2}. The inability of the algorithm to find distinct clusters corresponding to the four ground truth states highlights the high degree of overlap and the difficulty of the classification task.}
    \label{fig:cluster3}
\end{figure}

The high degree of overlap highlights the challenge of achieving perfect state identification. This is further illustrated in Figure~\ref{fig:cluster3}, where a standard k-means clustering algorithm (with k=4) is unable to correctly partition the synthetic data according to the four ground truth states. The gradual transitions between states, where the data is a mixture of two generative processes, further complicate the separation. This high degree of complexity makes our synthetic datasets a challenging and realistic benchmark for evaluating concept drift algorithms.

\section{Summary and Conclusions}
\label{sec:conclusion}

The prediction of stock market movements is challenging primarily due to the high amount of unknowns in the market's behavior (its efficiency), or in other words, its hidden context. This results in the non-stationary and adaptive nature of the acquirable information regarding financial markets. In this paper, we have addressed a critical bottleneck in the development of algorithms for this domain: the lack of a ground truth for evaluating performance in the presence of concept drift. Our primary contribution is a novel framework, namely ProteuS, for generating semi-synthetic financial time series with known, controllable structural breaks.

Using ProteuS, we have fitted ARMA-GARCH models to real-world ETF data and were able to capture the distinct statistical properties of different market regimes. We then simulated realistic, gradual, and abrupt transitions between these regimes, creating a rich and challenging set of datasets where the timing and nature of every concept drift are known. The subsequent feature engineering process, based on a comprehensive set of technical indicators, resulted in a high-dimensional data stream suitable for the evaluation of machine learning models in future research.

Our analysis of the generated datasets revealed a high degree of overlap between the different market states, confirming that the task of identifying and adapting to these regimes is non-trivial. This inherent difficulty makes our synthetic data a valuable resource for the research community, providing a robust benchmark for the development and validation of new algorithms for concept drift detection and adaptation.

While we aim to provide a valuable research tool, the successful application of any findings to real-world trading remains a complex endeavor. The high level of noise and the influence of external factors, such as news sentiment and fundamental economic data, mean that technical analysis alone is unlikely to yield a consistently profitable trading strategy. Future work should also focus on integrating these additional data sources into ProteuS and developing adaptive models to navigate the complexities of the financial landscape. Nevertheless, the ability to rigorously evaluate algorithms in a controlled, yet realistic, environment is a crucial step forward in the pursuit of more adaptive and intelligent financial forecasting systems.

\section*{Acknowledgements}
\label{sec:acknowledgements}
% We would like to thank the editor and the external reviewers for their thoughtful and detailed comments. 
Andres L. Suarez-Cetrulo wants to thank the European Commission for the funding under the Horizon Europe programme MANOLO (Grant Agreement No.101135782). David Quintana acknowledges financial support under grant ID2023-149827OB-C22 funded by MICIU/AEI/10.13039/501100011033 and by ERDF/UE.

% \appendix

%% References
%%   \citep{key}          ==>>  [#]


% \clearpage

% \section*{References}

%\bibliographystyle{elsarticle-num-names-alphsort}
% \bibliographystyle{elsarticle-harv}
% \bibliographystyle{elsarticle-num-names}
\bibliographystyle{model5-names}

\bibliography{main}
%\addbibresource

%% Authors are advized to submit their bibtex database files. They are





% \end{thebibliography}

\end{document}